\documentclass[12pt,preprint]{aastex}
\usepackage{multirow}
 \newcommand{\beq}{\begin{equation}}
 \newcommand{\eeq}{\end{equation}}
 \newcommand{\beqn}{\begin{eqnarray}}
 \newcommand{\eeqn}{\end{eqnarray}}

\begin{document}
 
 \title{\bf{The Axis-Symmetric Ring Galaxies: AM 0053-353, AM 0147-350, 
                                              AM 1133-245, AM 1413-243,
                                              AM 2302-322, ARP 318, 
 and Head-On Penetrations}}
 
 \author{Yu-Ting Wu and Ing-Guey Jiang}
 
 \affil{
 {Department of Physics and Institute of Astronomy,}\\ 
 {National Tsing-Hua University, Hsin-Chu, Taiwan} 
 }
 
 \begin{abstract}
 Axis-symmetric ring systems can be identified from the new catalog of 
 collisional ring galaxies in Madore et al. (2009).  
 These are O-type-like
 collisional ring galaxies. Head-on collisions by dwarf galaxies
 moving along the symmetric axis were performed through 
 N-body simulations to address their origins.
 It was found that the simulations with smaller initial relative velocities 
 between two galaxies, or the cases with heavier dwarf galaxies, 
 could produce rings with higher density contrasts.
 There are more than one generation of rings in one collision and 
 the lifetime of any generation of rings is about one dynamical time.
 It was concluded that head-on penetrations could explain these 
O-type-like ring galaxies identified from 
the new catalog in Madore et al. (2009), 
and the simulated rings resembling the observational 
O-type-like collisional rings are those at the early 
stage of one of the ring-generations. 
 
 \end{abstract}
 
 
 \section{Introduction}
 
 The morphology of galaxies has always been an attractive subject
 and has been investigated intensively through observational and 
 theoretical approaches, due to 
 both the beautiful shapes and richness of the complicated components in galaxies.
 In addition to normal spiral, elliptical, and irregular galaxies, 
 those dominated by ring-like structures, i.e. ring galaxies, are 
 one of the most intriguing categories of peculiar galaxies.
 
 Ring galaxies, a peculiar class of galaxies, contain various ring-like 
 structures with or without clumps, nuclei, companions, or spokes.
 For example, the famous Cartwheel galaxy, first discovered by 
 Zwicky (1941), shows an outer ring, an inner ring, a nucleus and spokes 
 (Theys \& Spiegel 1976; Fosbury \& Hawarden 1977; Higdon 1995).
 According to the morphology of ring galaxies, Few \& Madore (1986) studied 
 69 ring galaxies in the southern hemisphere and developed a classification 
 scheme.
 The ring galaxies are classified into two main classes: O-type and P-type 
 galaxies.
 The O-type galaxies have a central nucleus and smooth regular ring, while 
 P-type systems often contain an offset nucleus and a knotty ring.
 
 As more and more observational data on ring galaxies have been obtained 
 (Arp \& Madore 1987; Bushouse \& Standford 1992; Marston \& Appleton 1995; 
 Elmegreen \& Elmegreen 2006; Madore et. al. 2009), 
 three dominant theories have been proposed to explain the formation 
 and evolution of ring galaxies: the collision scenario, the resonance 
 scenario and the accretion scenario.
 In the collision scenario suggested by Lynds \& Toomre (1976), 
 ring galaxies are formed after a head-on collision between an 
 intruder galaxy and a disk galaxy.
 The formation of the Cartwheel galaxy is thought to be a prototype 
 of this scenario.
 Furthermore, from the observational statistics by Few \& Madore (1986), 
 P-type galaxies have an excess of companions and can be considered as being 
 formed by the collision scenario.
 In the resonance scenario, ring-like patterns are formed by gas accumulation 
 at Lindblad resonances, which respond to external perturbations 
 such as a bar or an oval.
 O-type galaxies, with a central nucleus and no obvious companions, 
 are thought to experience the resonance formation process 
 (for instance, IC 4214 (Buta et. al. 1999)).
 The third scenario, accretion, was proposed to explain 
 the origin of the polar ring galaxies.
 The polar ring galaxies contain a host galaxy and outer 
 rings with gas and stars that orbit nearly perpendicular to the 
 plane of the host galaxy.
 This type of galaxy is believed to be formed when the material 
 from another galaxy or intergalactic medium is accreted onto the host galaxy.
 A possible prototype of this scenario is the formation of NGC 4650A 
 (Bournaud \& Combes 2003).
 
 On the other hand,
 Elmegreen \& Elmegreen (2006) investigated 24 high redshift 
 galaxies with rings or partial rings and fifteen bent chain galaxies in the 
 GEMS (Galaxy Evolution from Morphology and SEDs; Rix et. al. 2004) 
 and GOODS (Great Observatories Origins Deep Survey; 
 Giavalisco et. al. 2004) fields.
They found that several rings are symmetric with centered nuclei 
and no obvious companions. For example, 
one of interesting ring galaxies is COMBO-17 No.44999, which has a bright 
 knotty structure like P-type galaxies but does not have companions.
 The brightness at the center of the galaxy COMBO-17 No.44999 is comparable 
 to that of the ring, so the densities at these two regions are similar.
Due to the lack of companions, it might be assumed they were formed through 
resonances. However, it is unclear whether there 
is any mechanism of resonances that could form these interesting rings because: 
(1) there is no obvious non-axisymmetric structure, i.e. bar or spiral,
in these galaxies; (2) the galaxy COMBO-17 No.44999
actually has a P-type property, i.e. knotty structures; 
and (3) the densities at the ring regions are very high.

Therefore, the above controversial  
results in Elmegreen \& Elmegreen (2006) 
encourage the study of P-type-like O-type ring galaxies or 
O-type-like P-type ring galaxies.
In fact, the dynamical relations between
companions and main galaxies for P-type ring galaxies
might not always be obvious. Those ring-structure-driving companions
could have merged with the main galaxies and disappeared, and 
thus formed a P-type-like ring without companions.
In contrast, the galaxies with axisymmetric rings might not have
formed through the interactions with those obvious surrounding companions. 
They could have been formed through a head-on merger, in which  
the dwarf galaxy was destroyed and overlapped with the center of 
the main galaxy, forming an O-type ring.

 Recently, Madore et al. (2009) 
 provided a new catalog of collisional ring galaxies.
 Although these galaxies are classified as the P-type,
 we tried to find axis-symmetric ring systems with central nuclei
 in this 
 catalog. In this study, we considered them as O-type-like 
collisional ring galaxies
 and investigated their formation history.     
 Due to these systems being axis-symmetric systems with 
 non-offset central nuclei, head-on collisions by dwarf galaxies
 moving along the symmetric axis are considered to be standard models.
 The stellar components of dwarf galaxies will overlap with the 
 central nucleus when the systems are viewed along the symmetric axes.
 Those identified companions around the ring galaxy, as listed in
 the catalog, 
 might not be the main drivers to produce these 
 axis-symmetric rings. 
 Many interesting questions could be addressed for the above 
 O-type-like collisional ring galaxies. 
 Could the standard models for head-on collisions explain these observed
 systems? 
 How bright or strong are these observed rings? 
 When would rings start to form during the collisions? 
 How long could ring structures survive? 
 These were the main goals for investigation in this paper.

 In fact, as a first attack, 
 Struck (2010) used an analytic theory to study a number of systems
 in Madore (2009). None of the galaxies studied in that
paper were studied in this paper.
 In order to provide more realistic 
 models, N-body simulations were used to study symmetric rings. 
The main differences between the proposed model and the one 
in Struck (2010) are that:
(1) impulse approximation is used in Struck (2010) but not in this study;
(2) in this study, close encounters between the main galaxy and 
the dwarf galaxy
    could happen several times during one collisional event 
in the simulations, 
    but the analytic model cannot have this;
(3) the dwarf galaxy and the main galaxy actually gravitationally influence
    each other, but only the proposed model's simulations can show this self-consistently; 
and (4) the dwarf galaxy finally merges and becomes part of the main disc 
    galaxy, but the model in Struck (2010) does not consider the contribution
    from the dwarf's stellar part.

 This study also developed a procedure to quantify the characteristics of 
 rings, which is helpful for the study of 
the formation and evolution of rings. Through the comparison between 
 observational and simulation results, the age of rings could be estimated.

 The observational data of 
 this study's O-type-like collisional ring galaxies are shown in Section 2,
 and the details of the models are shown in Section 3.  In Section 4,
 the results of the collision simulations and 
 the formation and evolution of rings are presented. 
 The observation-simulation comparisons are presented in Section 5,
the simulation of effect of the disc's length scale is in Section 6, 
and conclusions are given in Section 7.
  
\section{Observations}
 
 Madore (2009) presented a new catalog of 127 collisional ring galaxies, 
 with plausible colliders identified around their central galaxy in the
 images.
 In order to study these systems, and in particular to 
 investigate the relation between ring formation 
 and the processes of galaxy-intruder collisions, 
 all of the systems in the catalog were examined.
 As a first step, this study focused on symmetric cases;
 therefore, systems with axis-symmetric rings were chosen. 
 To make it easier to study ring structures, 
 only systems where the identified colliders were not closely 
 connected with the main galaxies were chosen, leaving 13 systems to study. 
 Finally, only those with known redshifts were used as sample systems 
 in this paper. These final six systems are shown in Fig. 1.
 The images, redshifts and the spatial resolution 
 were all obtained from the NASA/IPAC Extragalactic Database 
 (http://nedwww.ipac.caltech.edu/).
 The redshifts are written in the upper left corner of each panel 
 and the thick horizontal lines 
 in the lower left corner represent 10 arcsec.
 The image data were taken in waveband HeII ($\lambda$ = 468 nm)
 using a UK Schmidt Telescope (UKST), 
 which is a 1.2 meter Schmidt telescope located at the 
 Siding Spring Observatory (SSO) in Australia.
 The UKST was operated by the Royal Observatory, Edinburgh, 
 from 1973 to 1988, 
 and became part of the Anglo-Australian Observatory (AAO) thereafter.
 It has carried out many survey projects, for instance, 
 the ESO/SERC Southern Sky Survey, 
 the H-alpha Survey of the Milky Way and the Magellanic Clouds, 
 and also 6dF (6-degree Field).
 
 For each image in Fig. 1, as only the relative flux contrast 
 was important for this study, the flux value, $F$, 
 in all pixels was linearly rescaled to be within 
 the interval [0,1], 
 by multiplying a factor 
 $(F-F_{min})/(F_{max}-F_{min})$, 
 where $F_{min}$ and $F_{max}$ 
 were the minimum and maximum flux values in a particular image.

 To determine the length scale of these ring galaxies, 
 the spatial resolution of each pixel, 
 kpc/pixel, was calculated by the small-angle formula, where the 
 angular size of each pixel was 1.7 arcsec.
 The distance to the galaxy, which was needed in the small-angle formula, 
 was determined by the relativistic equation for the Doppler shift 
 and the Hubble law 
 with the Hubble constant $H_0=73$ $km$ $s^{-1}$ $Mpc^{-1}$
 (Spergel et al. 2007).
 Thus, the distance to the galaxy, $d$, was derived as follows:
 \begin{equation}
 d=\frac{\left(z+1\right)^2-1}{\left(z+1\right)^2+1}\left(\frac{c}{H_0}\right),
 \label {eq-dhlaw}
 \end{equation}
 where $z$ is the redshift of the galaxy and $c$ is the speed of light.
 
 In order to obtain the exact ring profiles of the above six galaxies, 
 this study concentrated on the main regions with rings.
 Fig.2 shows the flux-surface-density (i.e., flux per unit area) 
 plots around the area with ring regions.
 The center of each ring galaxy was at the origin (0,0)
 and the visible extent of the rings were 
 within the boundaries of the plots.
 The flux-surface-density values were rescaled to be from 0 to 1, 
 using the same method as in Fig. 1.
 
 Using the numerical values of the plots shown in Fig. 2, 
 the surface density profile was produced as a 
 function of distance $R$.
 To do that, each pixel was assigned a space coordinate.
 The pixel where the center of each ring galaxy is located
 was set as the coordinate origin (0,0).
 Other pixels' coordinates were set accordingly,  
 for example, the coordinate of the 
 pixel just to the left of the central pixel would be (-1,0).
 Given that coordinate system, the distance $r$ from the origin to a particular
 pixel could be defined easily.
 The pixels belonging to the $i$th annulus
 would be those with $i \textless r \leq i+1$, where
 $i=0, 1, 2, ...$ and $r$ is the pixel's distance to the origin.
 Then, the flux-surface-density of all the pixels belong to 
 the $i$th annulus 
 would be summed as the total flux-surface-density $F_i$. 
 The flux-surface-density, $\Sigma(R)$, of each annulus was defined 
 as $F_i/N_i$, where $N_i$
 is the total number of pixels belonging to this annulus.
 This study calculated $\Sigma(R)$ for all annuli and rescaled it 
 to be within the [0,1] interval. The data points shown in Fig.3 give
 the resulting density profiles. 
 
 To obtain an analytic curve to fit the above surface density profile, 
 this study first integrated the three-dimensional mass density 
 profile of the stellar disc in Hernquist (1993) to 
 yield the surface density profile as:
 \begin{equation}
 \Sigma_{H}(R) = \frac{M_d}{2\pi h^2}\exp \left( {\frac{-R}{h}} \right).
 \label {eq-ihs}
 \end{equation}
 
 It was then modified to be:
 \begin{equation}
 \Sigma_f(R) = 
 \alpha \frac{M_d}{2\pi h^2}\exp \left( {\frac{- \beta R}{h}} \right),
 \label {eq-ff}
 \end{equation}
 where two additional fitting parameters, $\alpha$ and $\beta$, were introduced.
 The best-fitting parameter set $(\alpha, \beta)$ 
 was determined by minimizing $\chi^2$ (Wall \& 
 Jenkins 2003) as:
 \begin{equation}
 \chi^2 = \sum^k_{i=1}\frac{\left( Q_i-E_i \right) ^2}{E_i}
 \label {eq-wj}
 \end{equation}
 where $Q_i$ is the surface density at a radius $R_i$ and $E_i$ 
 is the $\Sigma_f(R_i)$ with the 
 fitting parameter set $(\alpha, \beta)$.
 
 In the left panels of Fig. 3, i.e. Fig. 3(a), (c) and (e), 
 the empty circles and 
 the filled circles 
 are the surface density profiles of the ring galaxies.
 The dot and the dash lines are the best fitting profiles 
 for the empty circles and the filled 
 circles, respectively.
 The best fitting parameter set $(\alpha, \beta)$ for the ring galaxies 
 in Fig. 3(a), (c) and 
 (e) were $(12.44, 0.67)$, $(12.55, 0.70)$, 
 $(12.17, 0.65)$, $(10.74, 0.61)$, 
 $(15.44, 0.92)$ and 
 $(10.65, 1.06)$, in sequence.
 
 The density contrast was defined as $\Delta\Sigma(R)/\Sigma_f(R)$, 
 where $\Delta\Sigma(R)$ is the difference between the surface density 
 and the best-fitting 
 profile at a radius R, i.e. 
 $\Delta\Sigma(R)\equiv \Sigma(R)-\Sigma_f(R)$.
 Fig.3(b), (d) and (f) show the density contrast 
 as a 
 function of radius $R$ for the ring galaxies.
 These three panels showed that the location of the inner 
 ring structure for these six ring 
 galaxies was from 4 kpc to 20 kpc, 
 and the density contrast of the inner ring was from 0.09 to 0.9.
  
 \section{The Model}
 
 In the simulations,
 the target and the intruder were assumed to be a disc galaxy 
 and a dwarf galaxy respectively,
 in order to investigate the response of collisions between 
 a target disc galaxy, consisting of the stellar disc 
 and the dark matter halo, and a less massive dwarf galaxy, 
 containing the dark matter halo and the stellar component. 
 In order to study the effect of the intruder's mass,  
 two dwarf galaxy models were set up, i.e. $DG_A$ and $DG_B$. 
 
 \subsection{The Units}
 
 In the simulations, the unit of length was 1 kpc, 
 the unit of mass was $10^{10} M_\odot$, 
 the unit of time was $9.8\times10^8$ years, and  
 the gravitational constant $G$ was 43007.1.
 
 \subsection{The Initial Setting}
 
 Both the stellar disc and the dark matter halo of the target galaxy 
 followed the profiles employed in Hernquist (1993).
 The density profile of the stellar disc was: 
 \begin{equation}
 \rho_{d}(R,z)=\frac{M_d}{4 \pi h^2 z_0} \exp(-R/h) {\rm sech}^2(\frac{z}{z_0}),
 \label {eq3-1}
 \end{equation}
  where $M_d$ is the disc mass, $\it h$ is the radial scale length and $\it z_0$ is the vertical scale length.
 Furthermore, the halo density profile was:
 \begin{equation}
 \rho_{h}(r)= \frac{M_h}{2 \pi ^{3/2}}\frac{\alpha}{r_t r^2_c}\frac{\exp(-r^2 / r^2 _t)}{\frac{r^2}{r^2_c}+1} ,
 \label {eq3-2}
 \end{equation}
 where $M_h$ is the halo mass, $r_t$ is the tidal radius, and $r_c$ is the core radius.
 The normalization constant $\alpha$ is defined as:
 \begin{equation}
 \alpha = \left \{ 1-\sqrt{\pi}q \exp(q^2)[1-{\rm erf}(q)]\right \}^{-1},
 \label {eq3-3}
 \end{equation}
 where $q = r_c / r_t$ and ${\rm erf}(q)$ 
 is the error function as a function of $q$.
 
 The intruder dwarf galaxy was comprised of the dark matter halo 
 and a stellar part with Plummer spheres 
 (Binney \& Tremaine 1987; Read et. al. 2006), as follows:
 \begin{equation}
 \rho_{\rm plum}(r)= 
 \frac{3M_p}{4 \pi a^{3}}\frac{1}{\left ( 1+ \frac{r^2}{a^2} \right )^{2/5}},
 \label {eq3-4}
 \end{equation}
 where $M_p$ and $a$ are the mass and scale length of each component.
 
 The initial positions of the particles could be easily determined 
 according to the above given density profiles. 
 The initial velocities were assigned to particles through 
 the methods described in Wu \& Jiang (2009).
 
 The simulations were carried out with the parallel 
 tree-code GADGET (Springel, Yoshida \& White 2001), 
 and the softening lengths 
 were set as 0.05 kpc, 0.03 kpc and 0.035 kpc 
 for the target disc galaxy, the dwarf galaxy $DG_A$ 
 and the dwarf galaxy $DG_B$, respectively.

 \subsection{The Model Galaxies}
 
 Three model galaxies were needed
 in the simulations, including the disc galaxy 
 and the dwarf galaxies $DG_A$ and $DG_B$.
The mass and scale of the disc galaxy were chosen to be those of a standard
spiral galaxy, such as the Milky Way, using the details 
from Hernquist (1993).
 Using the chosen units, for the disc galaxy, 
 the disc mass, $M_d$, was 5.6, the disc radial scale length, 
 ${\it h}$, was 3.5, the disc vertical scale length, ${\it z_0}$, was 0.7, 
 the halo mass, $M_h$, was 32.48, the halo core radius, $r_c$, was 3.5 
 and the halo tidal radius, $r_t$, was 35.0.
 The disc galaxy had a total of 340,000 particles, 
 i.e. 290,000 dark matter particles and 50,000 stellar particles in the disc.
 The above parameters' values are summarized in Table 1.
  
 
\begin{table}[h]   
\begin{center}
 \begin{tabular}{llccc}
 \hline
 \multicolumn{2}{l}{Dark Halo}&&& \\
 mass $M_h$ ($10^{10} M_{\odot}$) & 32.48 &&\\
 core radius $r_c$ (kpc) & 3.5 &&\\
 tidal radius $r_t$ (kpc) & 35.0 &&\\
 number of particles & 290,000 &&\\
 &&&& \\
 \multicolumn{2}{l}{Stellar Disc}&&&  \\
 mass $M_d$ ($10^{10} M_{\odot}$) & 5.6 && \\
 radial scale length $h$ (kpc) & 3.5 &&\\
 vertical scale length $z_0$ (kpc) & 0.7 &&\\
 number of particles & 50,000 &&\\
 \hline
 \end{tabular}
\caption[Model parameters of the disc galaxy]{
Model parameters of the disc galaxy.
}
 \end{center}
  \end{table}

 Once the above parameters were set, 
 the dynamical time, $T_{dyn}$, could be defined by the velocity, $v_{1/2}$, 
 of a test particle at a disc's half-mass radius, 
 $R_{1/2}=5.95$, that is 
 $T_{dyn} \equiv 2 \pi R_{1/2}/ v_{1/2}=0.174$, where $v_{1/2}=214.77$.

 This study combined the dark matter halo and the stellar disc to set up 
 the disc galaxy, after constructing the components independently. 
 The components of the disc galaxy influenced each other and then approached 
 a new equilibrium.
 According to the virial theorem, when the disc galaxy is 
 in equilibrium, the value of $2K/|U|$ should be around one, 
 where $K$ and $U$ are total kinetic energy and total potential energy, 
 respectively.
 Thus, the same method as in Wu \& Jiang (2009) could be used 
 to examine the equilibrium of the disc galaxy.
 The disc galaxy approached a new equilibrium at $t = 15 T_{dyn}$ 
 and the energy conservation 
 was fulfilled because the total energy variation was 0.082$\%$.
 Hence, the disc galaxy at $t = 15 T_{dyn}$ was used to represent 
 the target disc galaxy at the beginning of the collision simulations.
 
 In order to study the effect of the mass ratio between 
 the main galaxy and intruder dwarf galaxy, 
 two dwarf galaxies with different masses, $DG_A$ and $DG_B$ were set up.
 For $DG_A$, the total mass was 9.52, which was a quarter of the mass 
 of the disc galaxy, and which had a mass to light ratio of 5.
 The scale lengths of the dark matter halo and the stellar component 
 were 3.0 and 1.5, respectively.
 The total number of particles was 85,000, containing 68,000 dark matter particles and 17,000 stellar particles.
 For $DG_B$, the total mass was 4.76, which was half of the mass of $DG_A$.
 The mass to light ratio, the scale lengths of the dark matter halo 
 and the stellar component was the same as in $DG_A$.
 To make sure the mass of each particle in the simulations was the same, 
 the total number of particles of $DG_B$ was 42,500, 
 including 34,000 dark matter particles and 8,500 stellar particles.
 Because the dark matter halo and the stellar component of the dwarf galaxy 
 were both spherically symmetric and could be set up simultaneously, 
 the whole dwarf galaxy with two components was initially in equilibrium.
 The parameters of $DG_A$ and $DG_B$ are summarized in Table 2.
 
 
\begin{table}[h]  
\begin{center}
 \begin{tabular}{lllcll} 
 \hline
       & \multicolumn{2}{c}{$DG_A$} && \multicolumn{2}{c}{$DG_B$}\\
 \cline{2-3} \cline{5-6} 
 & Dark Halo & Stellar Part && Dark Halo & Stellar Part \\
 \hline
 mass $M_p$ ($10^{10} M_{\odot}$) & 7.616 &1.904 && 3.808 & 0.952 \\ 
 scale length $a$ (kpc) & 3.0 &1.5 && 3.0 & 1.5\\
 number of particles & 68,000 & 17,000 && 34,000 & 8,500\\
 \hline
 \end{tabular}
\caption[Model parameters of dwarf galaxies]{
Model parameters of dwarf galaxies.
}
 \end{center}
  \end{table}

 \section{The Collision Simulations}
 
 The motivation of the present study was to examine 
 whether a collision scenario could account for 
 the origin of the axis-symmetric ring galaxies, 
 such as those described in Section 2.
 One target disc galaxy and one dwarf galaxy  
 were employed in each collision simulation.
 For each simulation, the stellar disc of the target galaxy lay on the $x-y$ plane
 and the initial relative velocity of the galaxies was along the $z$-axis.
 The initial separation of the disc galaxy 
 and the dwarf galaxy was 200 kpc, 
 which was far enough to make sure the two galaxies were initially well separated.
 
 In order to understand the effects of the mass ratio 
 and the initial relative velocity 
 between the target disc galaxy and the intruder dwarf galaxy,
 this study presented four collision simulations, which were the 
 combinations
 of two different initial relative velocity and mass ratios. 
 Table 3 lists the details of the four simulations, entitled S1, S2, S3, and S4.
 
\begin{table}[h] 
\begin{center}
 \begin{tabular}{cccc}
 \hline
 Simulation & Intruder & Initial Relative Velocity\\
 \hline
 S1 & $DG_A$ & 143.1 km/s\\
 S2 & $DG_A$ & 286.2 km/s\\
 S3 & $DG_B$ & 135.7 km/s\\
 S4 & $DG_B$ & 271.4 km/s\\
 \hline
 \end{tabular}
\caption[The details of four simulations]{
Intruders and initial relative velocities used in four simulations S1-S4.
}
 \end{center}
 \end{table}

 The initial relative velocities in S1 and S3 were derived 
 from parabolic orbits. Thus, the initial relative velocity, $v_i$, 
 was given according to the equation $E_{orb} = 0$,
 in which $E_{orb}$ is defined as: 
 \begin{equation}
 E_{orb} \equiv \frac{1}{2}\frac{M_1M_2}{M_1+M_2}v^2 _i-\frac{GM_1 M_2}{r_i},
 \label {eq4-1}
 \end{equation}
 where $r_i$ is the initial separation of two galaxies, 
 and $M_1$ and $M_2$ are the masses of the disc galaxy 
 and the dwarf galaxy, respectively.
 The initial relative velocity in S2 (S4) was simply 
 two times the one in S1 (S3), 
 therefore, S2 (S4) had a hyperbolic orbit.

 \subsection{The Evolution}
 
 This section presents both the orbital and morphological evolution 
 of the galaxies during the collisions. 
 To have more detail about the evolution of the galaxies in the simulations, 
 the time interval between each snapshot was set as $T_s = 0.1 T_{dyn}$.
  
 Fig.\,\ref{fig4}(a), (b), (c), (d) show the centers of mass
 of the disc and dwarf galaxies as a function of time
 during the collisions in S1-S4.
 The solid and dotted curves represent the centers of mass 
 of whole disc galaxy and dwarf galaxy 
 (both the stellar part and dark matter are included);
 the short dash and long dash curves are for the centers 
 of mass of the stellar disc and the dwarf's stellar component.
 For S1, i.e., in Fig.\,\ref{fig4}(a), the disc galaxy 
 and the dwarf ($DG_A$) had a close encounter at $t = 55T_{s}$ 
 and became well separated at $t = 65T_{s}$.
 After $10T_{s}$, i.e., at $t = 75T_{s}$, two galaxies became 
 close again due to the gravitational attraction.
 However, in S2, due to the higher initial relative velocity,
 the first close encounter took place earlier, the galaxies had 
 a much larger separation, and returned back at a later time,
 as shown in Fig.\,\ref{fig4}(b).
 Because the $DG_B$'s mass was only a half of $DG_A$'s,
 in S3 (Fig.\,\ref{fig4}(c)), 
 the gravitational force between the galaxies was smaller, and everything happened
 slightly later than in S1. 
 Finally, for S4 (Fig.\,\ref{fig4}(d)), 
 due to a weaker gravitational force and larger initial 
 relative velocity, the two galaxies became much more separated and did not return
 back as much as in previous cases. 
 
 In order to visualize the structure evolution, S1 was used as an example  
 to show the time 
 evolution of the stellar disc, together with the dwarf's stellar component.
 Fig.\,\ref{fig5} is the surface number density of the stellar particles 
 (the target and dwarf galaxies were both included) on the 
 $x-y$ plane, and Fig.\,\ref{fig6} is for the $x-z$ plane.
 The two galaxies were separated by 200 kpc initially,
 and approached each other from $t = 0$ to $t = 54T_{s}$ 
 during the first stage. 
 They made contact at $t = 55T_{s}$.
 Later on, the gravitational impact from the dwarf galaxy 
 warped the stellar disc upward and downward 
 from $t = 56T_{s}$ to $t = 75T_{s}$, as seen in Fig.\,\ref{fig6}.
 During this stage, the ring structure formed and expanded 
 outward, as shown in Fig.\,\ref{fig5}.
 Moreover, the dwarf galaxy started to expand after the encounter,
 and many of the dwarf galaxy's particles 
 escaped.
 Consequently, the stellar disc took on a layered appearance 
 at $t = 80T_{s}$ as viewed on the $x-z$ plane, 
 and its thickness increased with time.
 Most of the dwarf galaxy's particles were concentrated around the stellar disc, 
 but some of them extended to a distance of about 100 kpc.

 \subsection{The Ring}
 
 The results shown in Fig.\,\ref{fig5} and Fig.\,\ref{fig6} 
 indicated which snapshots contained ring structures, 
 therefore these snapshots were focused on. 
 Fig.\,\ref{fig7}(a) shows the surface density 
 of the stellar particles on the $x-y$ plane, 
 and also the fitted surface density profile 
 at $t=56T_s$, $t=57T_s$ and $t=58T_s$ in S1. 
 To obtain an analytic curve to fit the surface density profiles, 
 the modified surface density profile in eq.\,(\ref{eq-ff}) was used.
 The best-fitting parameter set $(\alpha, \beta)$ was also determined 
 by minimizing $\chi^2$, as in eq.\,(\ref{eq-wj}).
 In the $\chi^2$ fitting procedure, the surface density 
 beyond the radius $R_{end}$, where the surface density $Q_{end}$ is zero, is neglected.
 In Fig.\,\ref{fig7}(a), the open circles, 
 filled circles, and crosses represent the surface density 
 at $t=56T_s$, $t=57T_s$ and $t=58T_s$, respectively.
 The solid, dotted and short-dashed lines are the best-fitting 
 profiles with the parameter set 
 $(\alpha, \beta)$ = (2.5, 1.3), (2.3, 1.2) and (2.5, 1.2) 
 for open circles, filled circles and crosses, respectively.
 As shown in Fig.\,\ref{fig7}(a), it was clear that after the encounter, 
 a ring-like feature was evident at $t=56T_s$, 
 which then propagated outward as an expanding ring after $t=56T_s$.

 In order to determine the position of the ring, 
 the density contrast as a function $R$ was needed. 
 The $\Delta\Sigma(R)$ and the density contrast, 
 $\Delta\Sigma(R)/\Sigma_f(R)$ (as defined in Section 2), 
 as a function of radius $R$ at $t=56T_s$, $t=57T_s$ and $t=58T_s$ 
 are shown in Fig.\,\ref{fig7}(b) and (c), respectively.
 The different symbols (i.e. the open circles, filled circles and crosses) 
 represent the $\Delta\Sigma(R)$ and 
 the density contrast at different times 
 (as described in Fig.\,\ref{fig7}(a)).
 
 From the density contrast shown in Fig.\,\ref{fig7}(c), 
 the "ring region" around a ring-like feature could be determined
 to be the region in which all the density contrasts were larger than zero.
 Then, the average of density contrasts in this region, 
 $(\Delta \Sigma/\Sigma_f)_{av}$, 
 could be calculated and the corresponding radius $a$ and $b$, 
 where the density contrasts were equal to  $(\Delta \Sigma/\Sigma_f)_{av}$,
 could be determined. 
 The boundaries of the ring were then defined to 
 be at radius $a$ and $b$.
 The ring location, $L$, was defined by $L = (a+b)/2$ and 
 the width of the ring was $W = |b-a|$.
 
 Fig.\,\ref{fig7}(d) shows the number of particles as a function 
 of the angle in the ring region on the $x-y$ plane.
 The bin size of the angles was one degree. 
 The average number of particles in each bin was about 40, and the
 standard deviation was about 10. 
 Because there were no particular directions where 
 the number of particles was much larger or much less 
 than the average value,
 this panel confirmed that there were no big clumps, and 
 the shoulder-like features appearing 
 in the surface density profiles in Fig.\,\ref{fig7}(a) 
 were really the ring structures.
 
 To summarize, Fig.\,\ref{fig7} gives an example of the procedure 
 to determine the properties of ring structures.
 After fitting the surface density profile of all 
 stellar particles, 
 the density contrast as a function of $R$ 
 is calculated and used 
 to determine the boundaries, width, and the location of the ring.
 Finally, the numbers of particles in different directions on the $x-y$ plane 
 are checked, in order to make sure whether big clumps exist or not.
 This standard procedure was used to investigate the ring structures 
 in all of the simulation results.
 
 The density contrast as a function of $R$ at $t=71T_s$ of S1, 
 $t=51T_s$ of S2, $t=71T_s$ in S3, and also $t=47T_s$ in S4 
 are shown in Fig.\,\ref{fig8}(a)-(d).
 These were the snapshots when the first generation rings moved to
 the farthest distances in the simulations. The density contrasts 
 could be larger at this stage. 
 In addition, two peculiar density-contrast plots 
 at $t=89T_s$ and $t=91T_s$ in S3 are shown in
 Fig.\,\ref{fig8}(e)-(f).
 Fig.\,\ref{fig8}(e) shows that 
 two rings were formed around R=11 kpc and 17 kpc, individually.
 Because the inner ring, which was located at R=11 kpc,
 moved faster than the outer ring,
 these two rings merged together, as shown 
 in Fig.\,\ref{fig8}(f).
 
 Through the standard procedure to investigate ring structures, 
 the characteristics of rings at different times 
 in S1-S4 were shown, as illustrated in Fig.\,\ref{fig9}-Fig.\,\ref{fig12}, respectively.
 The average density contrast, location, and width of the ring 
 as a function of time are shown in Panel (a)-(c).
 The average number of particles in angular bins 
 (with a bin size of one degree) of the ring as a function of time
 is shown in Panel (d), in which the error bars
 are the standard deviations among different angular bins.
 
 Considering S1 in Fig.\,\ref{fig9} as the first example, 
 Panel (b) shows that the first ring was formed
 at $t = 55 T_s$, and this ring moved outward with a nearly constant 
 velocity until about $t = 70 T_s$. 
 In fact, there were three generations of rings. The 2nd was formed
 at $t = 77 T_s$ and the 3rd was formed at $t = 104 T_s$.
 As shown in Panel (a), the average density contrast 
 increased while the ring moved outward.
 Moreover, in a comparison of Panel (a) and (c), the width of 
 the ring was wider when the density contrast was higher. 
 Lastly, Panel (d) confirmed that the ring structures 
 at different times were smooth and without large clumps.
 
 For S2, Fig.\,\ref{fig10}(b) shows that 
 three generations of rings were formed at $t = 35 T_s$, 
 $t = 59 T_s$ and $t = 105 T_s$.
 The largest average density contrast was around five, and 
 the largest ring width was about 10 kpc, 
 as shown in Fig.\,\ref{fig10}(a) and (c).
 
 In S3, at particular times, i.e. from $t = 86 T_s$ to $t = 89 T_s$, 
 and also $t = 123 T_s$, two rings existed simultaneously.
 The open circles and crosses 
 in Fig.\,\ref{fig11} represent the corresponding values 
 of the additional ring. 
 The density contrast of the rings at $t = 89 T_s$ 
 was previously illustrated in Fig.\,\ref{fig8}(e).
 After $t = 89 T_s$, the two rings combined together and 
 formed a ring-like structure at $t = 90 T_s$.
 Finally, there is no cross symbol in Panel (c) 
 because the widths of the two rings at $t = 123 T_s$ 
 were the same, i.e., 2.8 kpc.
 
 Fig.\,\ref{fig12} shows the characteristics of  
 the ring-like structures in S4.
 In addition to the characteristics of the rings as described above, 
 it seems a ring-like structure existed at $R=6$ kpc
 from $t=86 T_s$ to $t=163 T_s$. This ring was fixed around 
 $R=6$ kpc, without moving outward.
 For this fixed ring-like feature, the average 
 density contrast was around 0.1 and the width was about 1 kpc.
 However, as shown in Panel (d), the huge deviations from 
 $t=86 T_s$ to $t=163 T_s$
 indicated that 
 this ring-like feature was not a ring but was clumps.
 
 In order to understand the effect of the initial 
 relative velocity between galaxies,
 this study compared the results between S1 and S2.
 Because the initial relative velocity in S2 was larger,
 the first generation ring in S2 formed earlier. 
 The higher average density contrast in S1 might have been due to the longer
 interaction timescale between galaxies, which was from the smaller 
 initial relative velocity.
 However, the location of the farthest ring and the width of the widest ring 
 were not closely related with the initial relative velocity. 
 Similar conclusions were obtained through the 
 comparison between the results of S3 and S4.
 
 On the other hand, comparing S1 with S3 (or S2 with S4)
 found that as the intruder dwarf galaxy was heavier, 
 the density contrast of the ring was higher 
 and the location of the farthest ring was further out.
 Finally, because the rings' moving velocities were close to the constants, 
 as shown in the plots of the ring locations as functions of time, 
 the rings' radial moving velocities could easily be calculated.
 The out-moving velocities of the first generation rings 
 in S1-S4 were 116.19 km/sec, 108.10 km/sec, 91.13 km/sec 
 and 86.75 km/sec, in sequence.
 
 \section{O-Type-Like Collisional Ring Galaxies}
 
 Having done the above simulations, this study next examined
 all of the snapshots in S1-S4 and checked which of them could resemble
 an O-type-like collisional ring galaxy.
 All of the simulations were set up as 
 pure axis-symmetrical simulations, and they did not produce any 
 non-axis-symmetric features, which could exist 
 in the galactic nucleus of observational images.

In order to compare the observational axis-symmetric ring galaxies 
with the simulation results, 
it was necessary to numerically determine the exact location of the ring 
for each observational image.
However, due to the poor observational resolution,
there were not enough data points available.
The location of the observational ring, $P$, 
was directly determined as the radius where the density contrast in the 
ring region was highest, 
as illustrated in Fig.\,\ref{fig3}.


In order to obtain the best theoretical model from the simulations
for each observational galaxy, the surface-mass-density 
profiles $\Sigma_{M}(R_M)$ of all snapshots in all simulations were
compared with the 
observed flux-surface-density profile $\Sigma_{F}(R)$.
To make $\Sigma_{M}(R_M)$ more consistent with $\Sigma_{F}(R)$,
new length units of simulations and mass-flux converting factors 
were considered.

The process was as follows.
First, choose a value $lu$ and $R_u=lu \times R_M$, so that the 
simulation ring position is redefined by this new length unit 
(the value of $lu$ is allowed only if it makes the ring position of 
that simulation snapshot to be between 0.8 and 1.2 of the observed ring 
location).
Second, the new surface-mass-density as a function of $R_u$
i.e. $\Sigma_{NM}(R_u)$ under this new length unit  is determined
(the time unit becomes different accordingly). 
Third, choose a mass-flux converting factor $MF$
and  $\Sigma_{MF}(R)= MF\Sigma_{F}(R)$.
Fourth, repeat the above with different values of 
$lu$ and $MF$ until the 
square of the difference between $\Sigma_{MF}(R)$
and $\Sigma_{NM}(R_u)$ (including all contributions 
from the available $R$) is the smallest.  
In other words, the process will stop when the 
smallest root-mean-square of deviations 
$RMS= {\sqrt{\frac{1}{N}\Sigma_{i=1}^{N} dev(R_i)^2 }}$ is obtained, 
where $dev(R_i)= \Sigma_{MF}(R_i)-\Sigma_{NM}(R_i)$, 
$R_i$ is the radii with available observational data, and
$N$ is the total number of observational data for this galaxy.               

The final results of this fitting are shown in Table 4. 
For each galaxy, the best snapshot resembling
the observed image is listed in the 2nd column.
The corresponding  $lu$ and $MF$ are in the 3rd and 4th columns.
The time unit becomes different accordingly 
after employing $lu$ to change the length unit,
and the real time of that snapshot is listed in the 5th column.
Finally, the $RMS$ is in the 6th column.

\vskip 0.3truein

\begin{table}[h]
\begin{center}
\begin{tabular}{cccccc} 
\hline
Galaxy & Snapshot & $lu$ & $MF$ ($10^4$ M$_\odot$/flux-count) & Time (Gyr) & $RMS$ ($10^{7}$M$_\odot$ kpc$^{-2}$)\\
\hline	
AM 0053-353 & S4 t=49 & 1.84 & 4.77 & 2.09 & 0.59 \\
AM 0147-350 & S4 t=49 & 1.67 & 1.79 & 1.80 & 2.33\\
AM 1133-245 & S4 t=49 & 1.93 & 2.53 & 2.24 & 0.57\\
AM 1413-243 & S3 t=61 & 2.05 & 5.22 & 3.05 & 0.48\\
AM 2302-322 & S2 t=35 & 3.31 & 5.29 & 3.59 & 1.03\\
ARP 318        & S4 t=37 & 1.54 & 0.89 & 1.21 & 13.46\\
\hline
\end{tabular}
\caption[Snapshots resembling
the observational O-type-like collisional ring galaxies.]{
Snapshots resembling
the observational O-type-like collisional ring galaxies.}
\end{center}
\end{table} 

The best-fitting density profiles are  
shown in Fig.\,\ref{fig13}, 
where the solid curves with full circles are the observational data,
and the dotted curves with empty circles are the
simulations.
It was found that the simulations could 
produce general trends of these profiles for
the first five galaxies in  Fig.\,\ref{fig13}(a)-(e).
Particularly, the simulation and observational profiles
were very close for the galaxy AM 1413-243 as shown 
in  Fig.\,\ref{fig13}(d).
The six best snapshots listed in Table 4 are plotted in
Fig.\,\ref{fig14}.
 
The above results showed that the head-on penetrations 
in simulations S1-S4 could explain the first five 
O-type-like collisional ring galaxies found from Madore et al. (2009).
For the sixth one, ARP 318, the major difference
was that the central part was very big and extended to a much larger radius.
The simulation profile dropped and could not fit the 
observed flat feature. 
In next section, a simulation is introduced, in which 
the disc's length scale is doubled. The purpose was to see if 
the larger disc could produce a bigger central part with a flat 
density profile, as well as to study the effect 
of different initial conditions for the major galaxy.

\section{The Effect of the Disc's Length Scale}
 
In S1-S4, this study mainly explored the effects of 
mass ratios and relative velocities between two merging galaxies.
It would be interesting to check whether the ring evolution 
presented here 
depended on the initial conditions of the primary galaxy.
For example, for the work about galactic discs, 
Chakrabarti \& Blitz (2009) and Chakrabarti et al. (2011) found that
disturbances in outer gas discs can be used to characterize 
galactic satellites, and that these disturbances are not extremely sensitive 
to the assumed initial conditions of the primary galaxies.

In order to obtain a better model 
for the galaxy ARP 318, a primary galaxy
with a larger length scale is used in the simulation presented here.
The disc galaxy was set up using the same method as described in Section 3, 
except
that the radial scale length $h$ was 7 kpc, 
which was two times the radial scale length
of the main disc galaxy in S1-S4.
After an initial  relaxation of 15 dynamical time, 
this disc galaxy approached a new equilibrium and
was used as the target galaxy in the merging simulation.

Because S3 provided the best fit to one of the O-type-like 
collisional galaxies described in the last section, all 
the model setting and details in this simulation were the same as S3, except for the 
length scale of the primary galaxy. 
That is, dwarf galaxy model $DG_B$ was employed as the intruder,
the initial separation was 200 kpc,
and the initial relative velocity was
135.7 km/s.

The results of ring evolution are shown in Fig.\,\ref{fig15}.
It was found that the general 
evolution process was very similar with the one in S3,
as there were 
also three ring generations.
The locations and widths of the rings were also similar with
those in S3.
The main difference was that the density contrast 
was smaller in this simulation. 
This could be because the disc had a larger length scale, 
so the amplitudes of any density disturbance became smaller.

All snapshots of this simulation were also compared with the six
observational galaxies through the same procedure described in 
the last section. 
Finally, it was confirmed that none of the snapshots in this 
simulation provided a better fit to any of the six observational
galaxies, including ARP 318.

\section{Concluding Remarks}
 
 Motivated by the results in Elmegreen \& Elmegreen (2006),
 this study found six axis-symmetric rings with central nuclei from the new catalog of
 collisional ring galaxies in Madore et al.(2009), and presented their 
 structures, profiles, and density contrasts. They were O-type-like
 collisional ring galaxies, and their possible formation scenario
 was addressed in this paper.
 Head-on collisions by dwarf galaxies
 moving along the symmetric axis were considered, 
 and N-body simulations were used to investigate
 the evolutionary process.
 It was found that the simulations with smaller initial relative velocities 
 between two galaxies or the cases with heavier dwarf galaxies 
 could produce rings with higher density contrasts.
 Usually, there were more than one generation of rings 
 in any of the simulations.  
 The lifetime of any generation of rings was about one dynamical time,
 and the rings continued moving outward with constant velocities until they   
 disappeared at the outer part of the galactic discs.

 Through the observation-simulation comparison shown in  
Fig.\,\ref{fig13}, 
 this study concluded that head-on penetrations could explain these low
 density contrast ring galaxies. Moreover, it was found that 
the simulation rings that resemble the observational 
O-type-like collisional rings were those at the early 
stage of one of the ring-generations.

 \section*{Acknowledgment}
The authors acknowledge the anonymous referee for the very helpful
comments that improved this paper. 
The authors would like to thank the National Center for High-performance 
Computing
 for their computer time and facilities. 
 This work was supported in part 
 by the National Science Council, Taiwan, under 
 NSC 98-2112-M-007-006-MY2.

 \begin{figure}
 \includegraphics[angle=0,scale=.80]{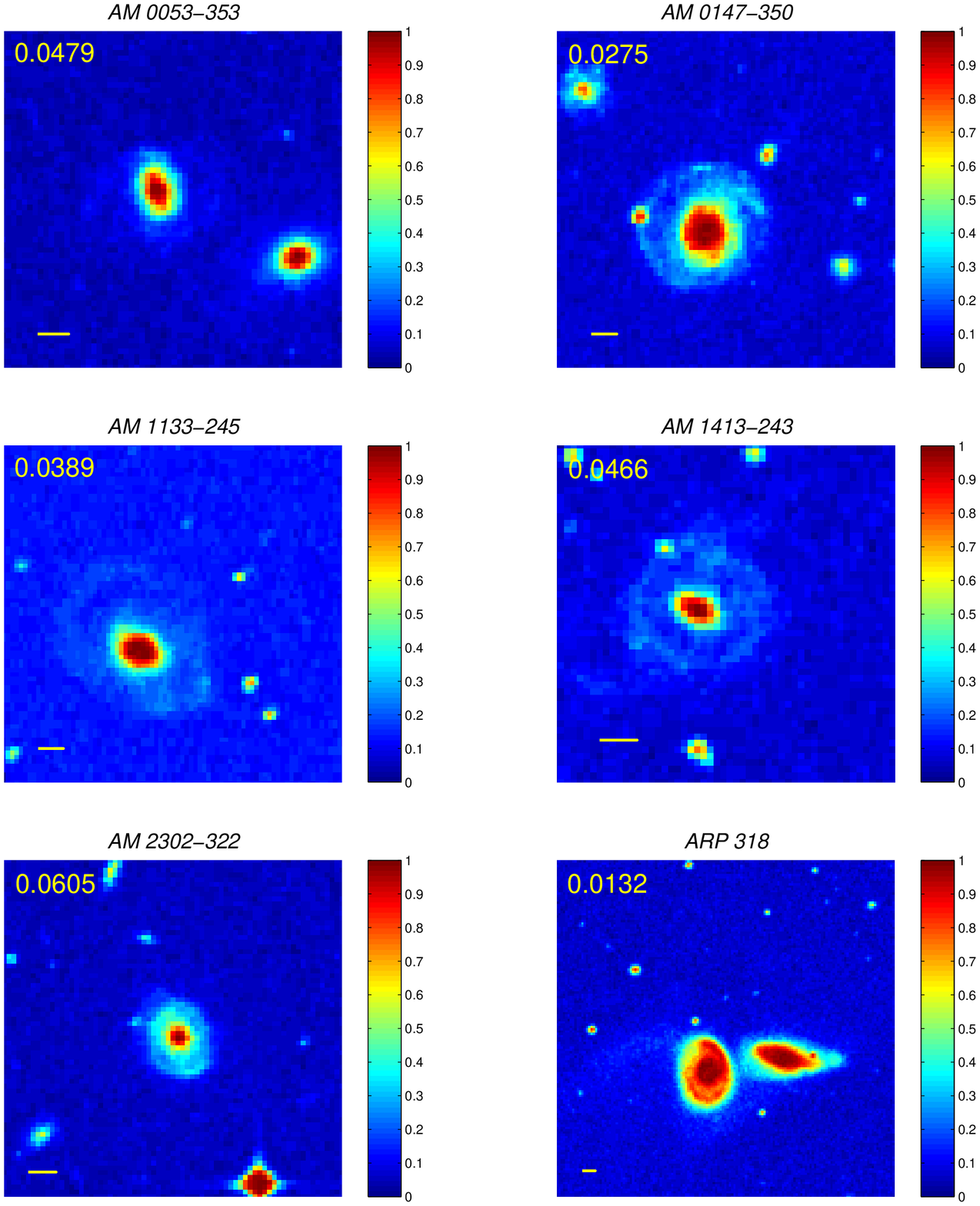}
  \caption{The images of six axis-symmetric ring galaxies.
 These images were obtained using UK Schmidt Telescope (UKST).
 Redshifts are in the upper left corner; thick horizontal lines represent 10 arcsec.
 }
 \label {fig1}
 \end{figure}
 
 \clearpage
 \begin{figure}
 \includegraphics[angle=0,scale=.80]{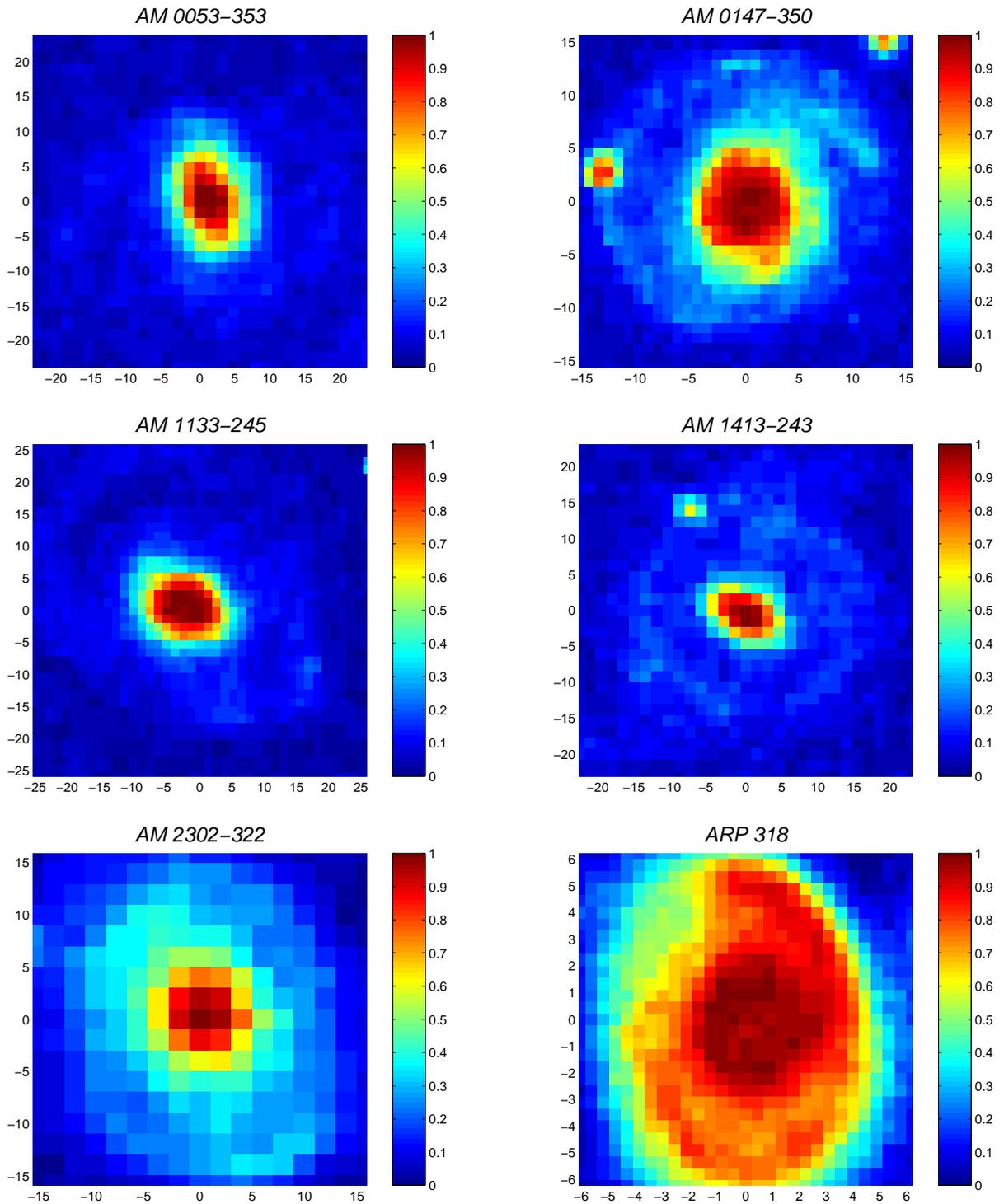}
  \caption{The flux-surface-density plots of ring galaxies around ring regions.
 The unit of length is 1 kpc.
 }
 \label {fig2}
 \end{figure}
 
 \clearpage
 \begin{figure}
 \begin{center}
 \includegraphics[angle=0,scale=.70]{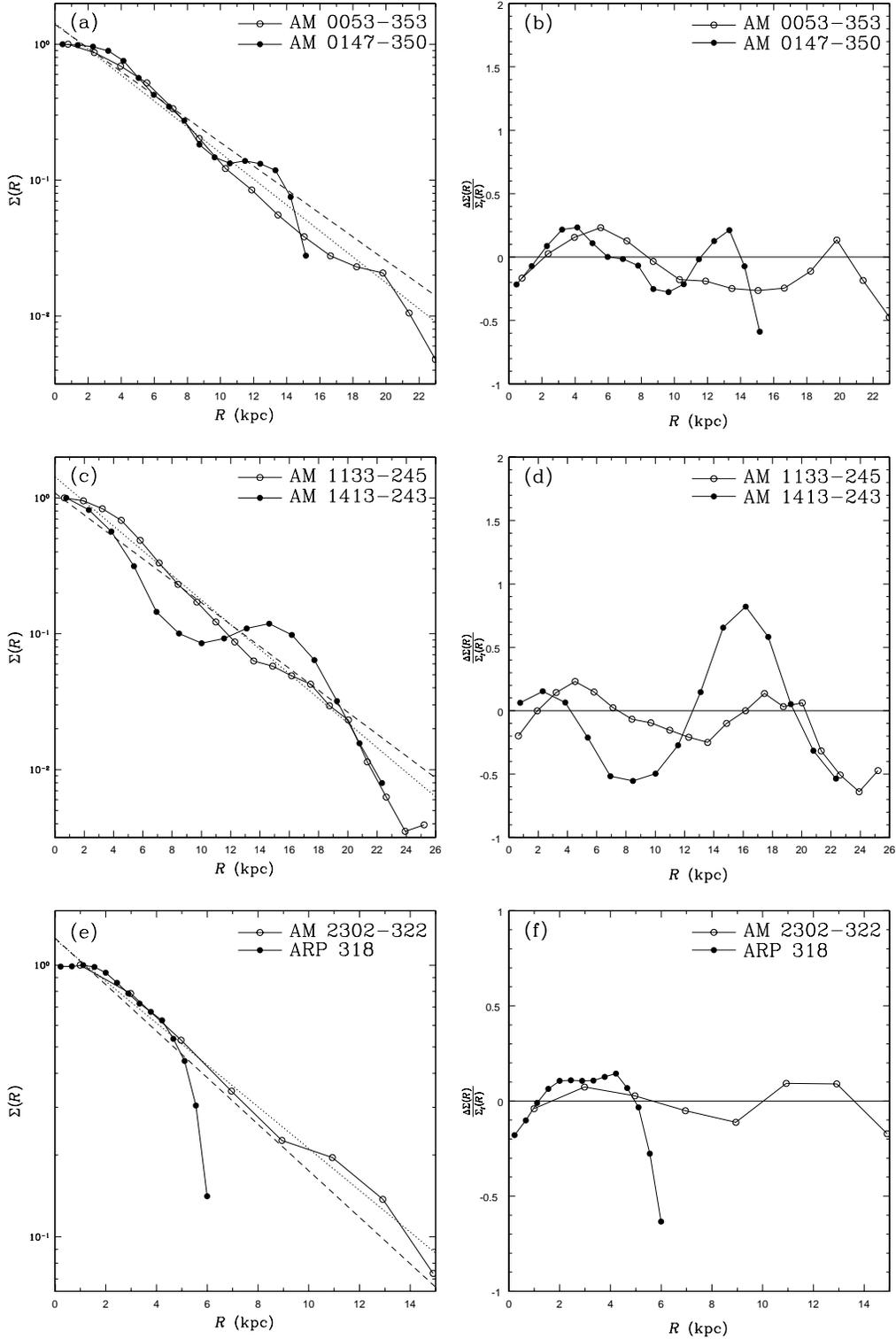}
  \caption{The surface density profiles and the density contrasts of ring galaxies.
 For panels (a), (c) and (e), the empty circles and the filled circles represent the surface density profiles of ring galaxies.
 The dot and the dash lines show the fits to the empty circles and the filled circles, respectively.
 For panel (b), (d) and (f), the empty circles and the filled circles show the density contrasts of ring galaxies.
 }
 \label {fig3}
 \end{center}
 \end{figure}

 \clearpage
 \begin{figure}
 \includegraphics[angle=0,scale=.80]{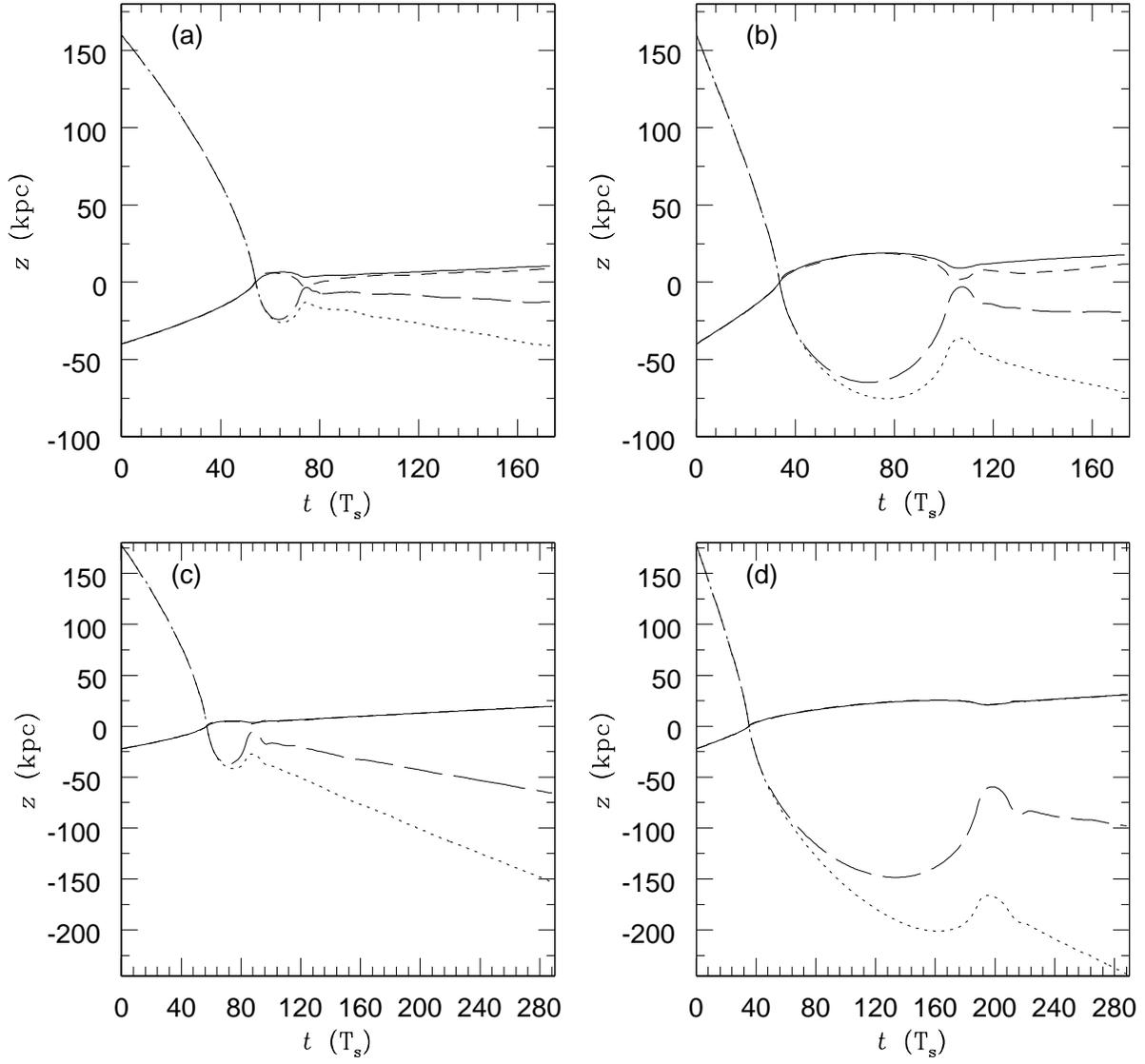}
  \caption{The evolution of S1-S4.
 Panel (a)-(d) show the distance between the dwarf galaxy and the disc galaxy as a function of time in S1-S4, respectively.
 The solid and dotted curves correspond to the centers of mass of whole disc galaxy and dwarf galaxy.
 The short dash and long dash curves represent the centers of mass of the stellar disc and the dwarf's stellar component.
 }
 \label {fig4}
 \end{figure}
 
 \clearpage
 \begin{figure}
 \begin{center}
 \includegraphics[angle=0,scale=.90]{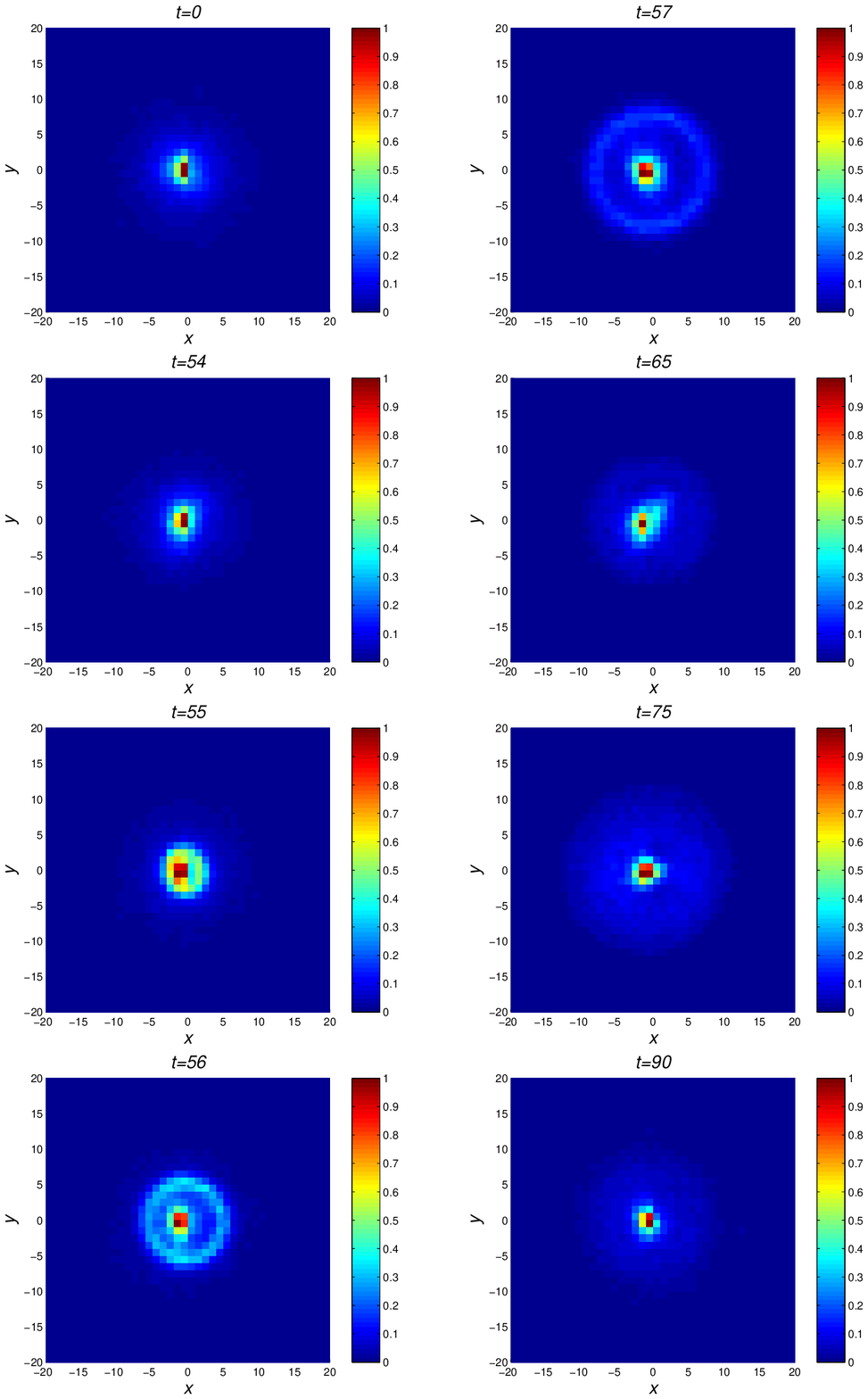}
  \caption{Time evolution of the projected density of stellar components, 
 including the stellar disc and the dwarf's stellar part, on the x-y plane in S1.
 }
 \label {fig5}
 \end{center}
 \end{figure}

 \begin{figure}
 \begin{center}
 \includegraphics[angle=0,scale=.90]{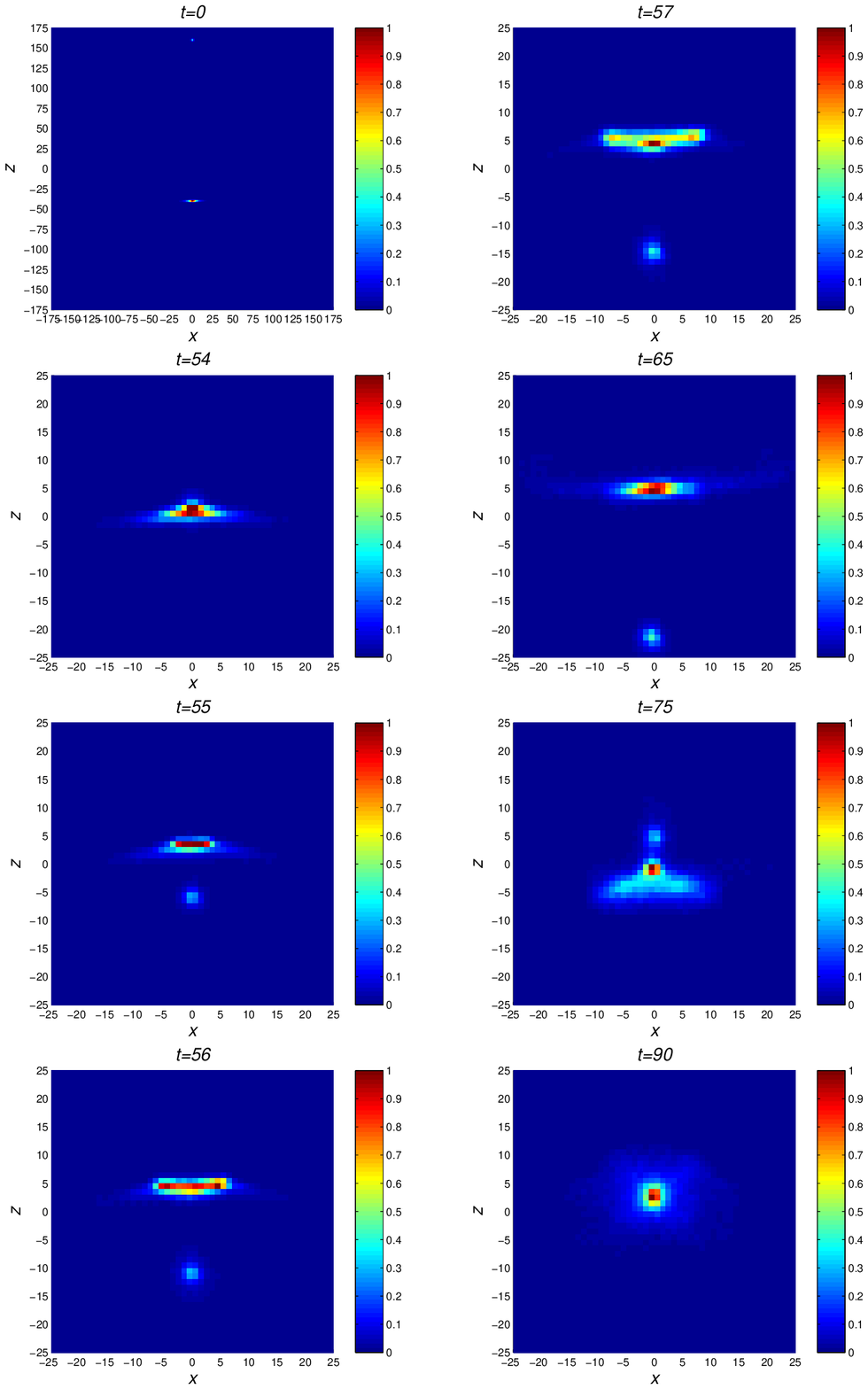}
  \caption{Time evolution of the projected density of stellar components, 
 including the stellar disc and the dwarf's stellar part, on the x-z plane in S1.
 }
 \label {fig6}
 \end{center}
 \end{figure}
 
 \clearpage
 \begin{figure}
 \includegraphics[angle=0,scale=.80]{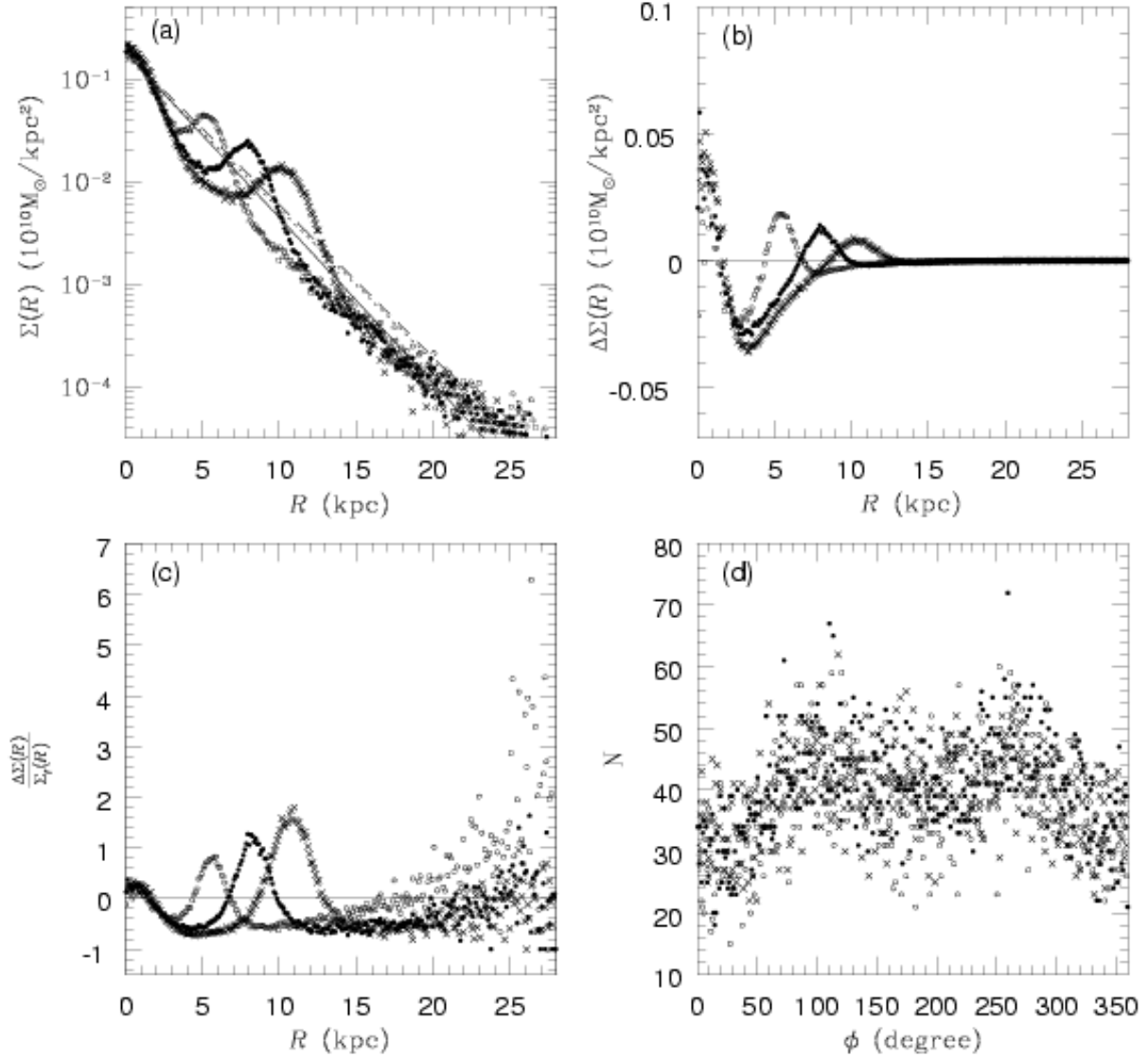}
  \caption{
 (a) The surface density profiles and the best-fitting profiles at the different times.
 The open circles, filled circles and crosses show the surface density profiles at $t=56T_s$, $t=57T_s$ and $t=58T_s$.
 The solid, dotted and short-dashed lines are the best-fitting profiles for the above three different kinds of points, respectively.
 (b) $\Delta\Sigma ({\it R})$, which is the difference between the surface density and the best-fitting profile, 
 as a function of radius at $t=56T_s$ (open circles), $t=57T_s$ (filled circles) and $t=58T_s$ (crosses).
 (c) The density contrast as a function of radius at $t=56T_s$ (open circles), $t=57T_s$ (filled circles) and $t=58T_s$ (crosses).
 (d) The number of particles within one degree as a function of angle in the ring region on the x-y plane 
 at $t=56T_s$ (open circles), $t=57T_s$ (filled circles) and $t=58T_s$ (crosses).
 }
 \label {fig7}
 \end{figure}
 
 \clearpage
 \begin{figure}
 \includegraphics[angle=0,scale=.90]{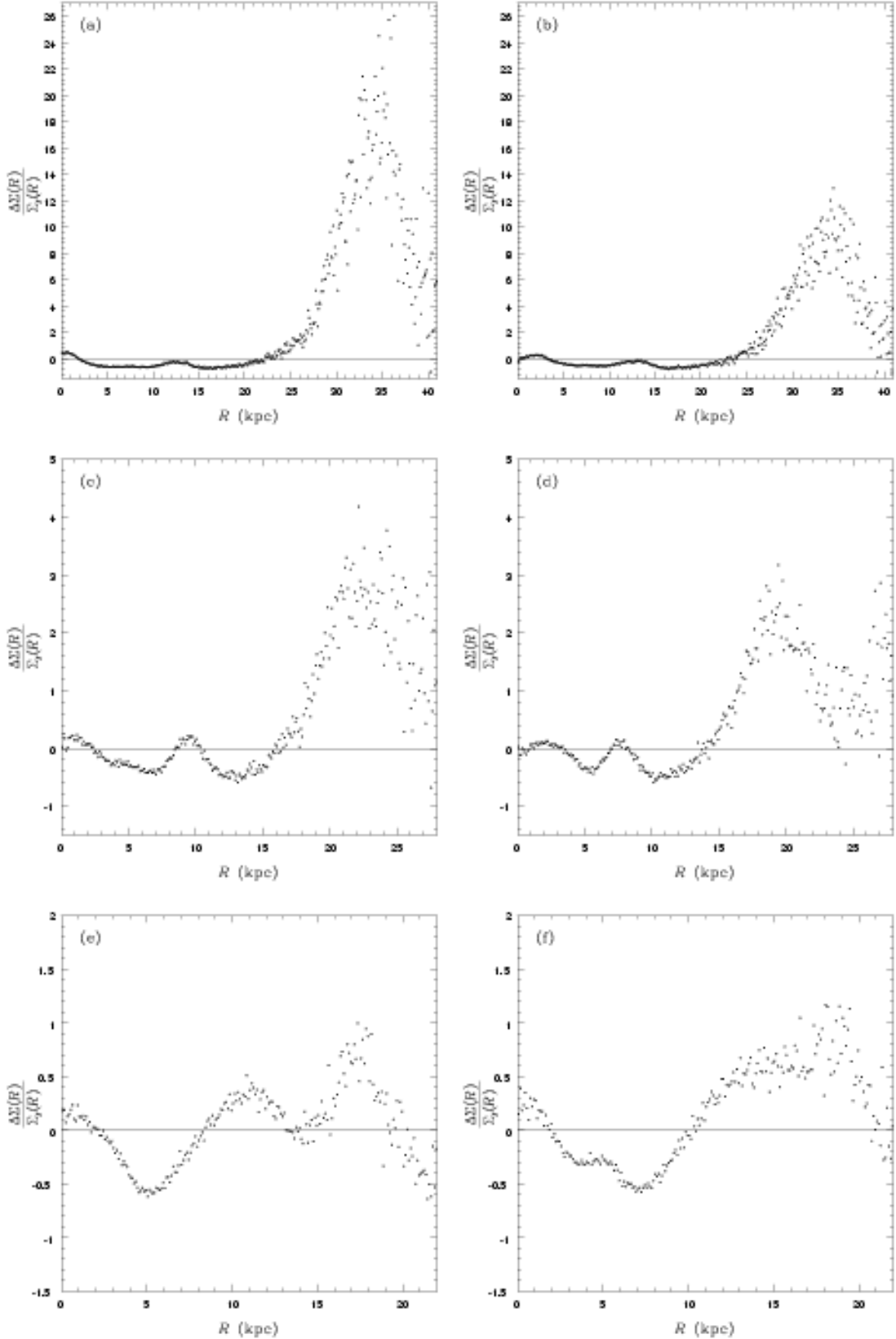}
  \caption{The density contrast as a function of radius.
 (a) $t=71T_s$ in S1; (b) $t=51T_s$ in S2; (c) $t=71T_s$ in S3; (d) $t=47T_s$ in S4; (e) $t=89T_s$ in S3; (f) $t=91T_s$ in S3.
 }
 \label {fig8}
 \end{figure}
 
 \clearpage
 \begin{figure}
 \includegraphics[angle=0,scale=.80]{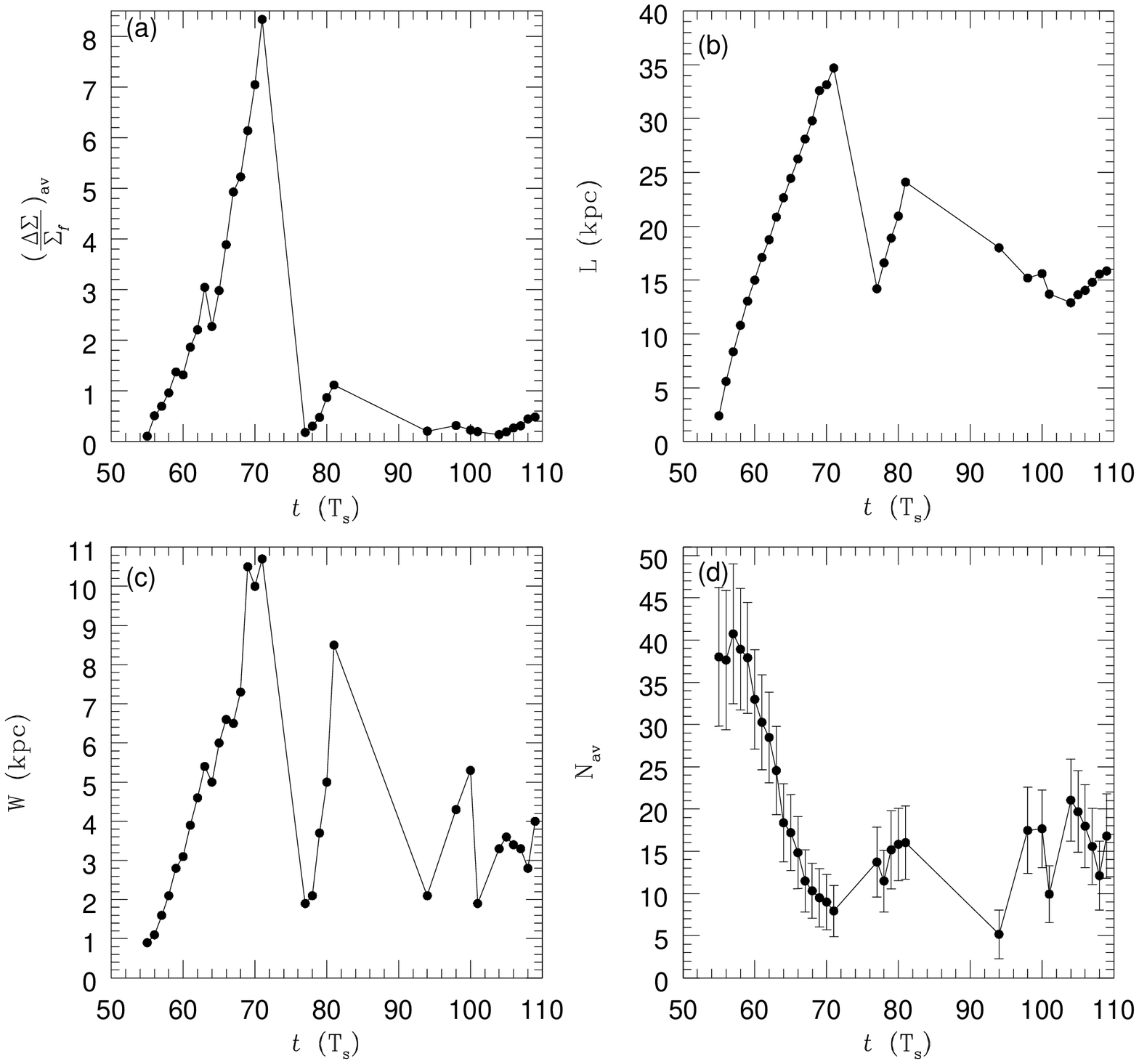}
  \caption{The characteristics of rings in S1.
 (a) The average of the density contrast in the ring region as a function of time.
 (b) The location of the ring as a function of time.
 (c) The width of the ring as a function of time.
 (d) The average number of particles in angular bins. 
 The error bar shows the variation in different angular bins.
 }
 \label {fig9}
 \end{figure}
 
 \clearpage
 \begin{figure}
 \includegraphics[angle=0,scale=.80]{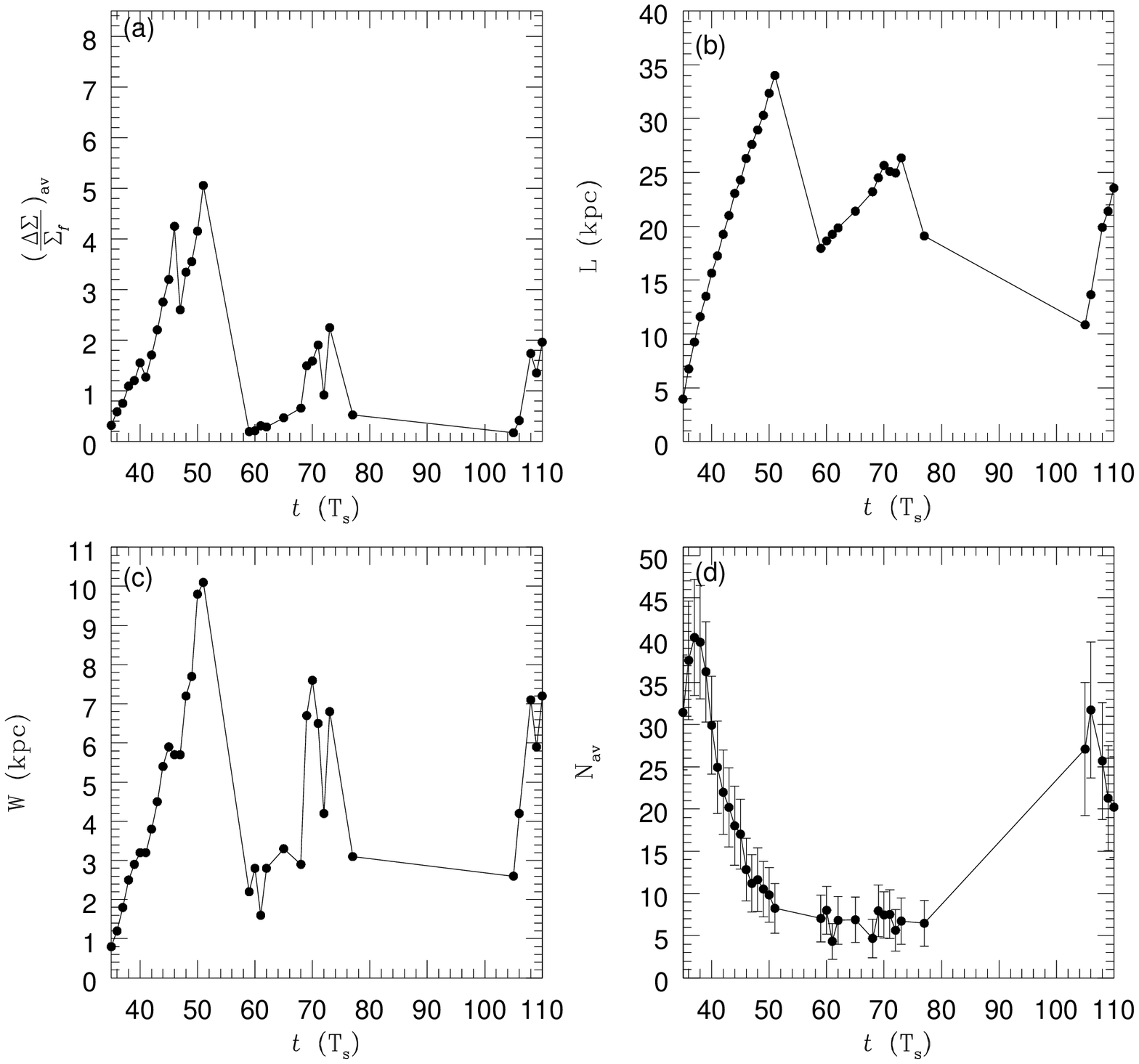}
  \caption{The characteristics of rings in S2.
 (a) The average of the density contrast in the ring region as a function of time.
 (b) The location of the ring as a function of time.
 (c) The width of the ring as a function of time.
 (d) The average number of particles in angular bins. 
 The error bar shows the variation in different angular bins.
 }
 \label {fig10}
 \end{figure}
 
 \clearpage
 \begin{figure}
 \includegraphics[angle=0,scale=.80]{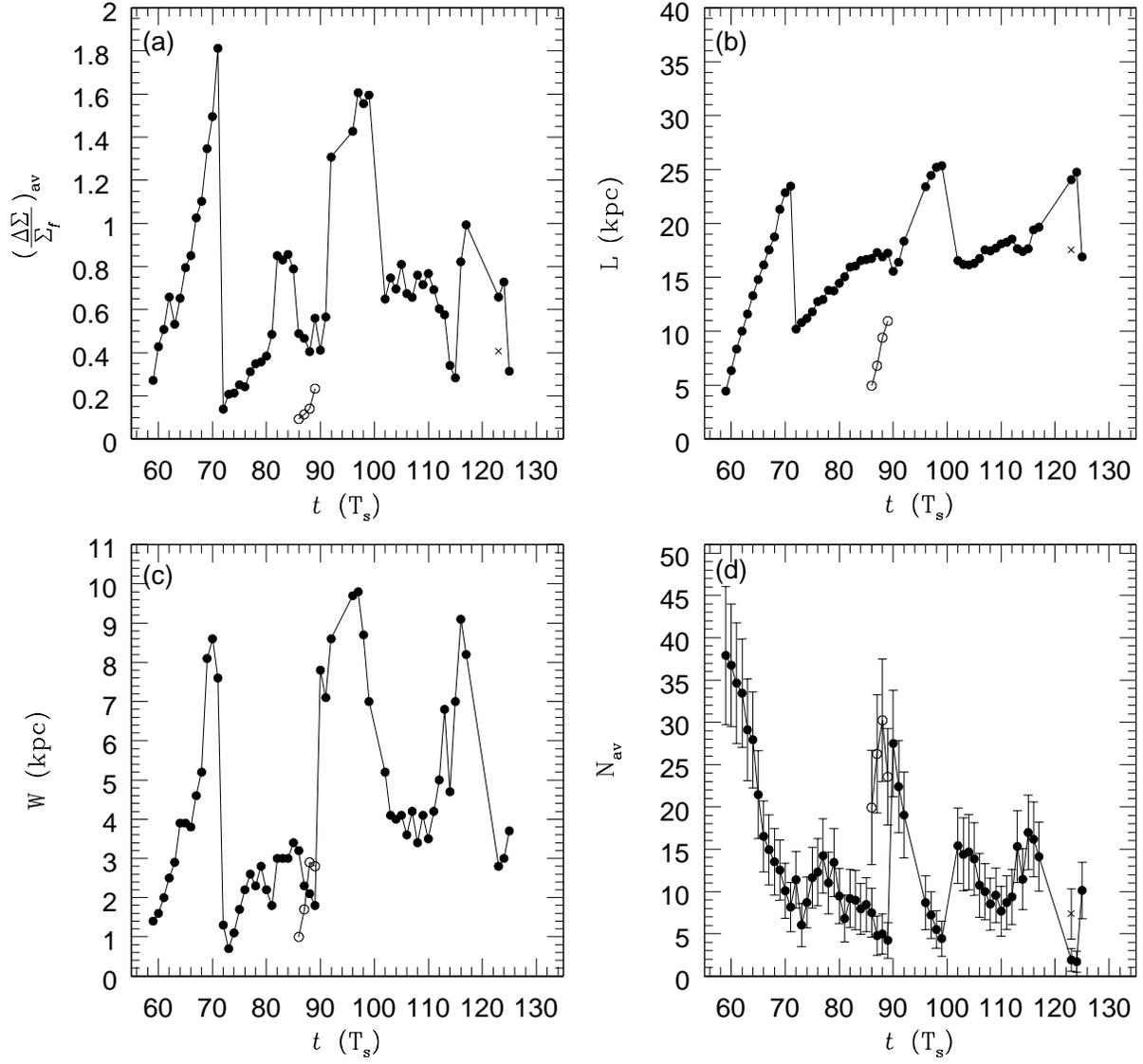}
  \caption{The characteristics of rings in S3.
 (a) The average of the density contrast in the ring region as a function of time.
 (b) The location of the ring as a function of time.
 (c) The width of the ring as a function of time.
 (d) The average number of particles in angular bins. 
 The error bar shows the variation in different angular bins.
 The open circles and crosses represent the additional ring at $t=86T_s$ - $89T_s$ and $t=105T_s$.
 }
 \label {fig11}
 \end{figure}
 
 \clearpage
 \begin{figure}
 \includegraphics[angle=0,scale=.80]{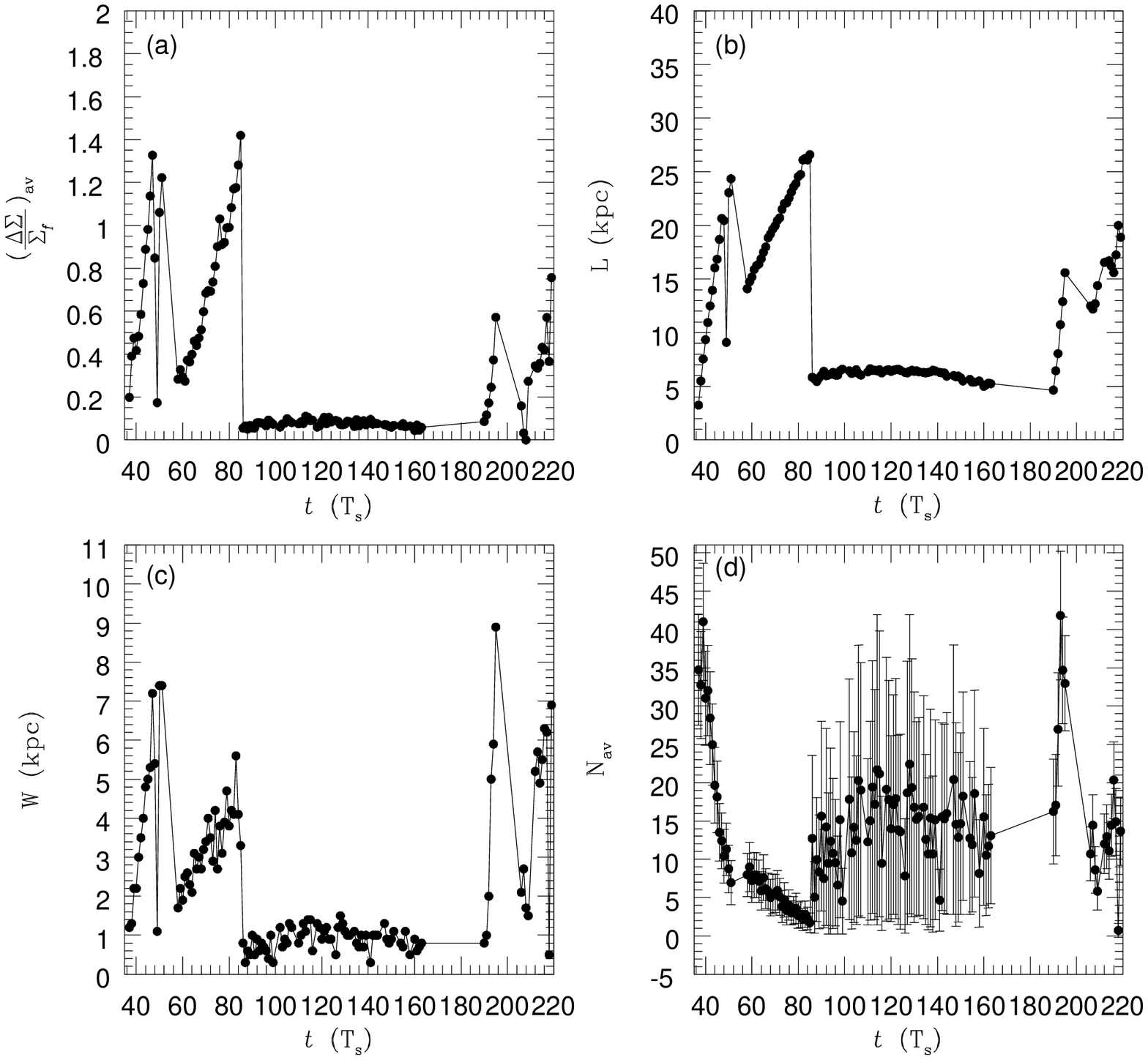}
  \caption{The characteristics of ring-like structures in S4.
 (a) The average of the density contrast in the ring region as a function of time.
 (b) The location of the ring as a function of time.
 (c) The width of the ring as a function of time.
 (d) The average number of particles in angular bins. 
 The error bar shows the variation in different angular bins.
 }
 \label {fig12}
 \end{figure}
 
  \clearpage
 \begin{figure}
 \includegraphics[angle=0,scale=.80]{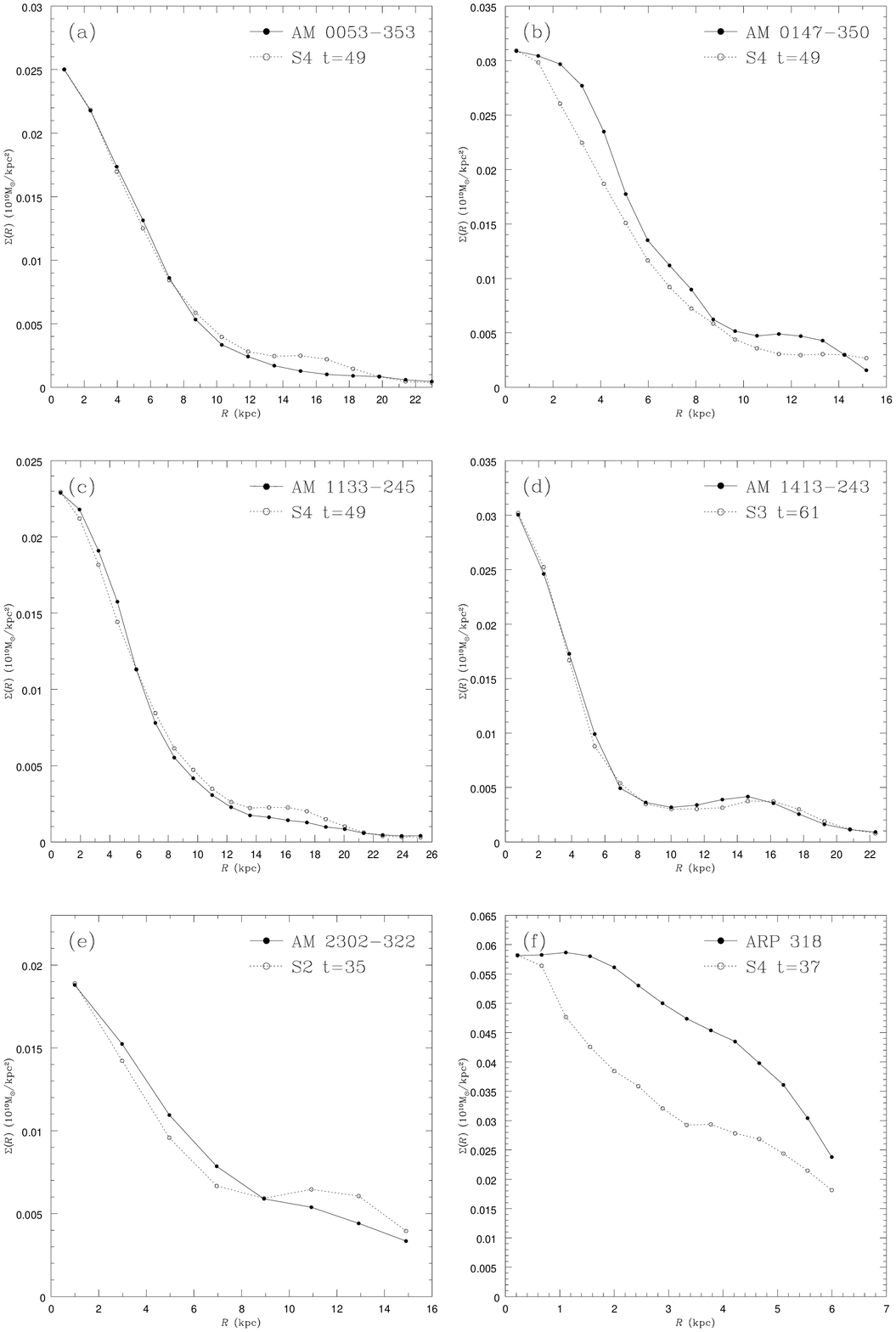}
 \caption{The radial density profiles of the 
observational ring galaxies and the corresponding
simulational snapshots resembling them.
 }
 \label {fig13}
 \end{figure}

  \clearpage
 \begin{figure}
 \includegraphics[angle=0,scale=.80]{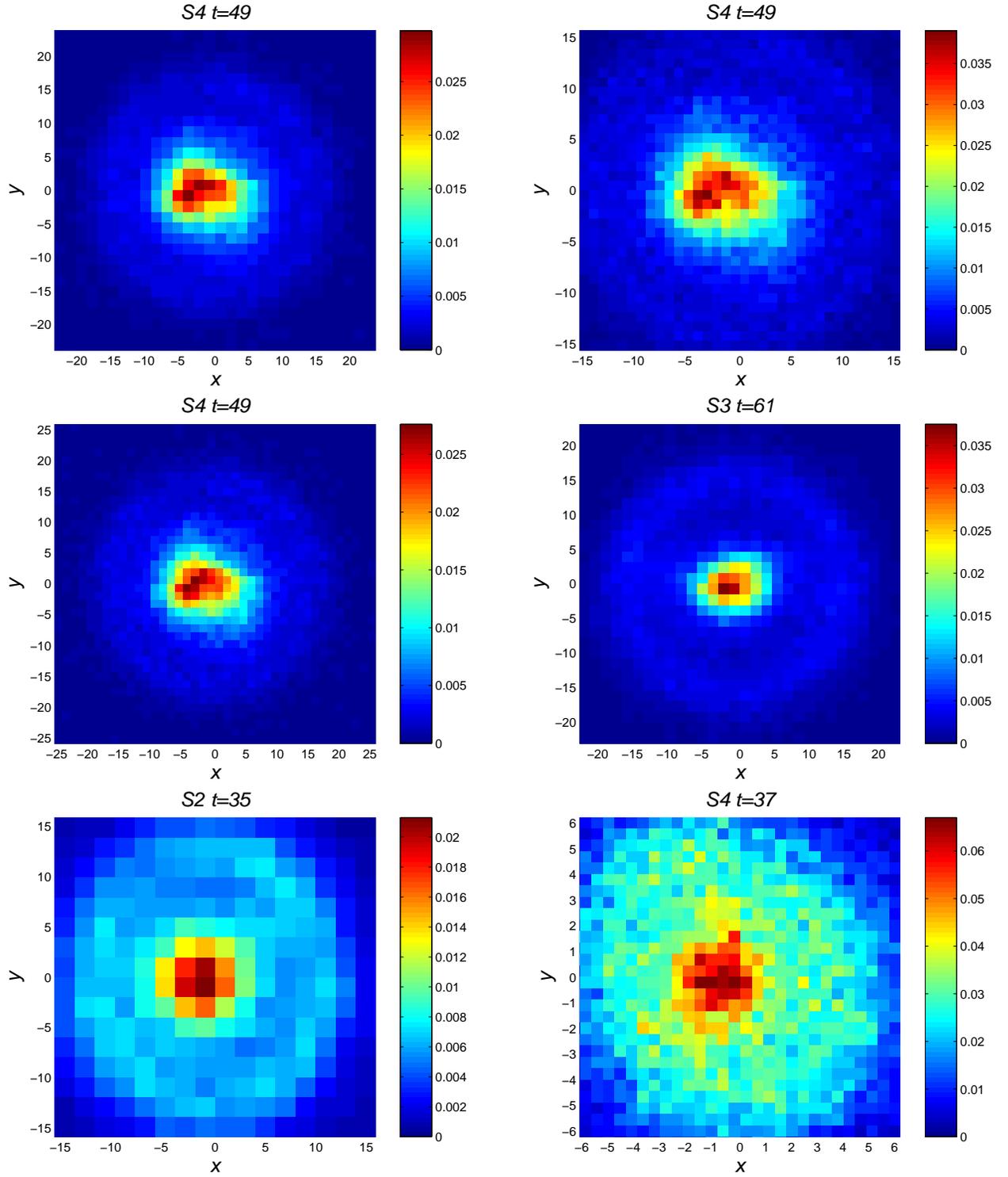}
 \caption{The simulational snapshots resembling 
the observational O-type-like collisional ring galaxies.
 }
 \label {fig14}
 \end{figure}

\clearpage
 \begin{figure}
 \includegraphics[angle=0,scale=.80]{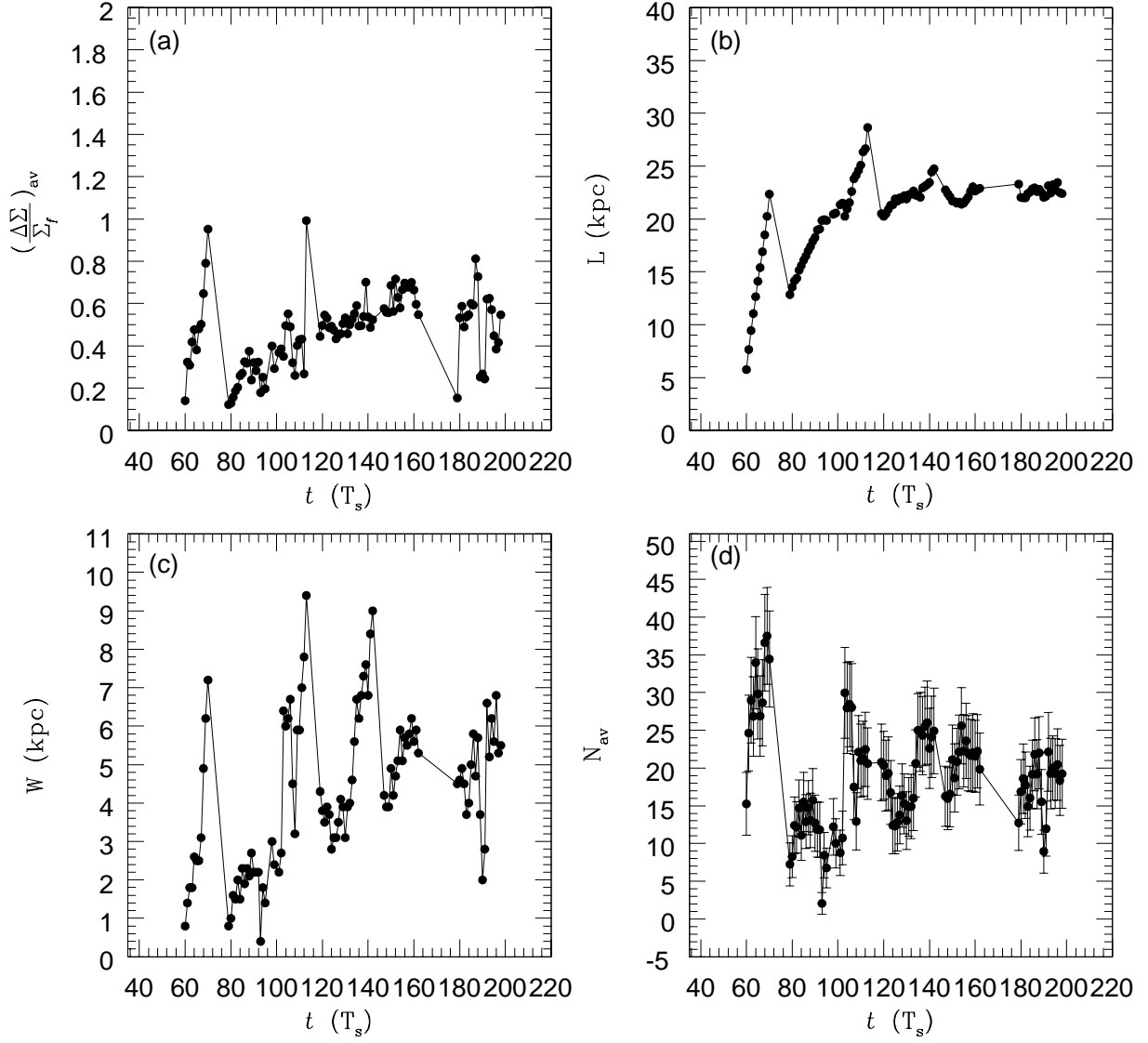}
  \caption{The characteristics of rings 
in the simulation described in Section 6.
 (a) The average of the density contrast in 
the ring region as a function of time.
 (b) The location of the ring as a function of time.
 (c) The width of the ring as a function of time.
 (d) The average number of particles in angular bins. 
 The error bar shows the variation in different angular bins.
 }
 \label {fig15}
 \end{figure}

 \end{document}